\documentclass[lettersize,journal]{IEEEtran}
\pdfoutput=1
\usepackage{amsmath,amsfonts}
\usepackage{algorithmic}
\usepackage{algorithm}
\usepackage{array}
\usepackage[caption=false,font=normalsize,labelfont=sf,textfont=sf]{subfig}
\usepackage{textcomp}
\usepackage{stfloats}
\usepackage{url}
\usepackage{verbatim}
\usepackage{graphicx}
\usepackage{cite}
\usepackage{amssymb} 
\usepackage{ntheorem} 

\newtheorem{theorem}{Theorem}
\newtheorem{lemma}{Lemma} 
\allowdisplaybreaks[4] 
\begin{document}

\title{Uplink Transmission Design for \\Fluid Antenna-Enabled  Multiuser MIMO Systems with Imperfect CSI}

\author{Linyue Hu, Luchu Li,  Cunhua Pan,~\IEEEmembership{Senior Member,~IEEE,} Hong Ren,~\IEEEmembership{Member,~IEEE}
\thanks{Linyue Hu is with the College of 
	Electronics and Information Engineering, Sichuan University, 
	Chengdu 610065, China. (e-mail: hulyue@163.com)}
\thanks{Luchu Li, Cunhua Pan and Hong Ren are with National Mobile Communications Research Laboratory, Southeast University, 
	Nanjing 211189, China. (e-mail: 220240860, cpan, hren@seu.edu.cn)} }



\maketitle

\begin{abstract}
This paper investigates a two-timescale uplink transmission framework for a fluid antenna-enabled multiuser multi-input multi-output system (MIMO-FAS). Antenna positions are optimized based on statistical channel state information (CSI), while beamforming vectors at the base station (BS) adapt to instantaneous CSI. Under a Rician fading channel with imperfect CSI, we establish a linear minimum mean square error (LMMSE)-based channel estimation approach and derive a closed-form expression for the achievable uplink rate using a low-complexity maximal-ratio-combining (MRC) detector. The optimization problem is formulated as a minimum user rate maximization problem by optimizing the fluid antenna positions, subject to the feasible region and the minimum spacing distance constraints. To address this non-convex problem, a genetic algorithm (GA) method is proposed, encoding antenna configurations as population individuals. Additionally, an accelerated gradient ascent algorithm is proposed to enhance computational efficiency. Numerical results validate the mathematical derivations and demonstrate that the proposed two-timescale transmission strategy significantly outperforms traditional FPA systems, with both algorithms achieving enhanced gains.

\end{abstract}

\begin{IEEEkeywords}
Fluid antenna system (FAS), two-timescale transmission design, antenna position optimization, MIMO, channel estimation.
\end{IEEEkeywords}

\section{Introduction} \label{sec1}
\IEEEPARstart{O}{ver} the past decades, wireless communication technologies have evolved in response to the escalating demands for low-latency and high-capacity services \cite{ref1}. As a cornerstone of modern wireless communication, multiple-input multiple-output (MIMO) provides enhanced spectral efficiency and reliability through beamforming gain, spatial multiplexing and diversity gains \cite{ref2}. To date, MIMO technology has developed rapidly, transitioning from single-user MIMO (SU-MIMO) to multiuser MIMO (MU-MIMO), and more recently to massive MIMO. Nevertheless, prohibitive hardware costs and energy consumption associated with large-scale antenna arrays impede sustainable deployment for massive MIMO \cite{ref3}, \cite{ref4}. Furthermore, conventional MIMO systems relying on fixed-position antennas (FPAs) suffer from static spatial configurations with limited degrees of freedom (DoFs) that fail to adapt to dynamic user distributions and time-varing channel conditions, thereby constraining the overall performance of the system.

To address these limitations, fluid antenna system (FAS), alternatively termed movable antenna (MA), has garnered considerable attention for its potential to overcome the challenge of FPAs. As a novel paradigm characterized by position and shape flexibility, FAS leverages advanced reconfigurable antenna technologies, including reconfigurable radio-frequency pixel arrays \cite{ref5}, liquid-based antennas \cite{ref6}, and motorized mechanical movable antennas \cite{ref8}. In contrast to traditional FPA systems, FAS allows for dynamic repositioning of antennas within a predefined region, introducing additional DoFs in the physical layer. 
Notably, \cite{ref9} has demonstrated that FAS can achieve spatial diversity with a single antenna, making it particularly suitable for size-constrained mobile devices. By enabling antennas' movement to cope with user distribution and channel conditions, FAS alleviates issues associated with FPAs that struggle to make full use of spatial diversity and multiplexing, thereby enhancing network capacity and reliability\cite{ref10}. Furthermore, the fluid antenna can be repositioned from interference-prone areas to locations with more favorable channel conditions, facilitating efficient interference management without the need for additional antennas \cite{ref11}. 

\IEEEpubidadjcol
Initial FAS investigations focused on single-input single-output (SISO) systems, establishing theoretical foundations through outage probability analysis of single-antenna FAS\cite{ref12} and demonstrating comparable capacity with multi-antenna systems \cite{ref13}.
Motivated by these works, researchers have devoted considerable efforts to studying FAS, ranging from performance analysis to channel estimation. \cite{ref14} provided a closed-form expression of the average level crossing rate in FAS, \cite{ref15} analyzed the FAS performance over general correlated channels, \cite{ref16} proposed a novel sequential linear minimum mean-squared error (LMMSE)-based channel estimation method for a small number of fluid antenna ports, \cite{ref17} studied the achievable performance of FAS when multiple ports could be active, and \cite{ref18} developed a switched combining (SDC) scheme for FAS to reduce channel estimation overhead. A recent review has summarized literature related to FAS \cite{ref19}. Several researches further indicated that the priority of hybrid architectures combining FAS with physical layer security \cite{ref20}, integrated sensing and communication (ISAC) \cite{ref21} \cite{ref22}, and reconfigurable intelligent surfaces (RISs) \cite{ref23}. 

Recent advancements further studied the FAS-enhanced MIMO (MIMO-FAS) system. To fully unleash the potential of MIMO-FAS, several works focused on antenna position optimization. \cite{ref24} explored the optimal diversity and multiplexing tradeoff of MIMO-FAS to reveal the fundamental limits by optimizing antenna position, which can also be viewed as optimizing the ports. Expanding on this concept, \cite{ref25} developed a new MA-enabled MIMO, which is effectively a MIMO-FAS, demonstrating that flexible antenna positioning enhances MIMO capacity by reconfiguring the channel between transceivers. Furthermore, the authors in \cite{ref26} investigated uplink transmission where each user is equipped with a single fluid antenna, and minimized the total transmit power of all users by jointly optimizing each user’s antenna position and the base station (BS) receive beamforming matrix. Further advancing the application of fluid antennas, \cite{ref27} proposed a new BS architecture employing multiple fluid antennas in uplink transmission. Additionally, \cite{ref28} jointly optimized transmitted power and antenna positions in a cell-free massive MIMO system with fluid antennas at access points to maximize the minimum uplink rate across users. At the same time, various optimization algorithms, including the alternating optimization (AO) algorithm in \cite{ref25}, \cite{ref28} and particle swarm optimization (PSO) algorithm in \cite{ref27}, have been proposed and validated their effectiveness. 

However, prevailing FAS studies rely critically on two fundamental yet impractical assumptions. The first is the availability of perfect instantaneous channel state information (CSI), while the second is the requirement for real-time adjustment of antenna positions. In practice, due to the fact that the frequent updates of antenna positions in each channel coherence time, the instantaneous CSI changes too rapidly to be reliably obtained, especially in fast-fading channel \cite{ref16}, \cite{ref29}. On the other hand, mechanical latency in fluid antenna relocation hinders its adaptability to dynamical channel variations, posing a challenge for FAS in fast-moving transceiver locations \cite{ref30}. 

To address these two practical challenges, some researchers considered the use of statistical CSI for antenna position optimization \cite{ref31}, \cite{ref32}, \cite{ref33}. In \cite{ref32}, the authors developed a rate maximization framework for transmit precoding and transmit/receive fluid antenna position designs with statistical CSI in the MIMO-FAS system. \cite{ref33} introduced a two-timescale downlink transmission framework for MIMO-FAS with perfect CSI. In the two-timescale scheme, the beamforming vectors are designed based on instantaneous CSI to counteract channel fluctuations. 
More importantly, the antenna positions are optimized based on the long-term CSI, such as the signal power, the locations, the angles of arrival (AoAs), and the angles of departure (AoDs), which vary much slower than the instantaneous CSI. The antenna positions only need to be updated when there is a change in large-scale channel information. Unlike instantaneous CSI-based schemes, which require to update the antenna positions at each channel coherence interval, the two-timescale transmission design reduces both computational complexity and energy consumption, while effectively handling rapid channel variations and mitigating their impact. Nonetheless, the work in \cite{ref33} only considered downlink transmission and assumed that perfect CSI is available. To deal with the difficulty of acquiring perfect CSI, \cite{ref34} demonstrated a novel CSI-free fluid antenna position optimization method, which treats position optimization as a derivative-free problem, utilizing zeroth-order gradient approximation techniques to adaptively adjust the position. In the context of imperfect CSI, the uplink two-timescale transmission scheme in MIMO-FAS remains an open issue.

To bridge this gap, this paper proposes a two-timescale uplink transmission scheme for a multiuser MIMO-FAS operating under a Rician fading channel with imperfect CSI. 
The optimization problem is inherently challenging due to the non-convexity of the objective function and constraints, as well as the requirement for expectation over random states. The main contributions of this work are summarized as follows:

\begin{enumerate}{}{}
	\item{First, we propose a two-timescale uplink transmission framework for multiuser MIMO-FAS that optimizes the antenna position based on statistical CSI, while optimizing BS's beamforming vectors based on instantaneous CSI. This hierarchical design significantly decreases fluid antenna reconfiguration frequency and channel estimation overhead.}
	\item{Next, we consider the Rician channel model with imperfect CSI to assess the impact of the line-of-sight (LoS) and non-line-of-sight (NLoS) channel components. To estimate the channel at the BS, we employ the LMMSE method. Using a low-complexity maximal-ratio-combining (MRC) detector, we derive a closed-form expression for the use-and-then-forget (UatF) bound of the uplink achievable rate. This expression is valid for any finite number of fluid antennas and only depends on the statistical CSI.}
	\item{We formulate an optimization problem to maximize the minimum user rate by optimizing the fluid antenna positions, subject to constraints on the feasible moving region for antennas and the minimum inter-antenna distance.}
	\item{To solve this non-convex problem efficiently, we first propose a genetic algorithm (GA)-based method to obtain a suboptimal solution. In this approach, each individual in the population represents a specific realization of fluid antenna positions and its fitness function combines minimum user rate with a penalty term for constraint enforcement.}
	\item{To further reduce the computational complexity, an accelerated gradient ascent-based algorithm is proposed to solve this fairness problem. This approach employs a log-sum-exp approximation to handle the non-differentiable minimum user rate maximization objective function, coupled with adaptive step size and projection operation to ensure antenna positions within the designated area.}
	\item{Numerical results demonstrate that the proposed two-timescale framework for MIMO-FAS achieves significant improvement over the traditional FPA system with imperfect CSI. It is observed that both the GA-based method and the accelerated gradient ascent algorithm provide enhanced performance, offering a balance between accuracy and computational efficiency.}
\end{enumerate}

The remainder of this paper is structured as follows. Section \ref{sec2} details the channel model and problem formulation. Section \ref{sec3} analyzes the LMMSE channel estimation method. Section \ref{sec4} derives a closed-form lower bound expression of the achievable rate. Section \ref{sec5} proposes a GA-based method and an accelerated gradient ascent algorithm for solving the minimum user rate maximization optimization problem. Section \ref{sec6} illustrates numerical results, followed by conclusions in Section \ref{sec7}.

\textit{Notations:}  $\mathcal{X}$, $\mathbf{X}$, $\mathbf{x}$ and x represent a set, a matrix, a vector, and a scalar, respectively. $(\cdot)^{*}$, $(\cdot)^{T}$, $(\cdot)^{H}$ and $(\cdot)^{-1}$ represent the conjugate, transpose, Hermitian transpose and inverse operation, respectively. $\mathbb{E} \{\cdot\}$ and $\text{Tr}\{\cdot\}$ denote the expectation and covariance operators. $[\mathbf{X}]_{m, n}$ denotes the $(m, n)$-th entry of the matrix. $\mathbb{R}^{M \times N}$ and $\mathbb{C}^{M \times N}$ are the space of $M \times N$ real and complex sets. $\mathbf{0}$ and $\mathbf{I}_N$ are the $N \times N$ identity matrix and null vector. For the vector, $\| \cdot \| $ denotes the Euclidean norm and $| \cdot |$ denotes the modules of the complex number. $| \mathcal{X} |$  represents the cardinality of the set $\mathcal{X}$. $\mathbf{x} \sim \mathcal{CN}(\bar{\mathbf{x}}, \mathbf{C})$ represents a complex  Gaussion distributed vector with mean vector $\bar{\mathbf{x}}$ and covariance matrix $ \mathbf{C}$. $x \sim \mathcal{U}[m, n]$ indicates that $x$ follows a uniform distribution within the range from $m$ to $n$. 

\section{System Model} \label{sec2}
\subsection{Multiuser MIMO-FAS with imperfect CSI}
\begin{figure}[!t]
	\centering
	\includegraphics[width=3.4in]{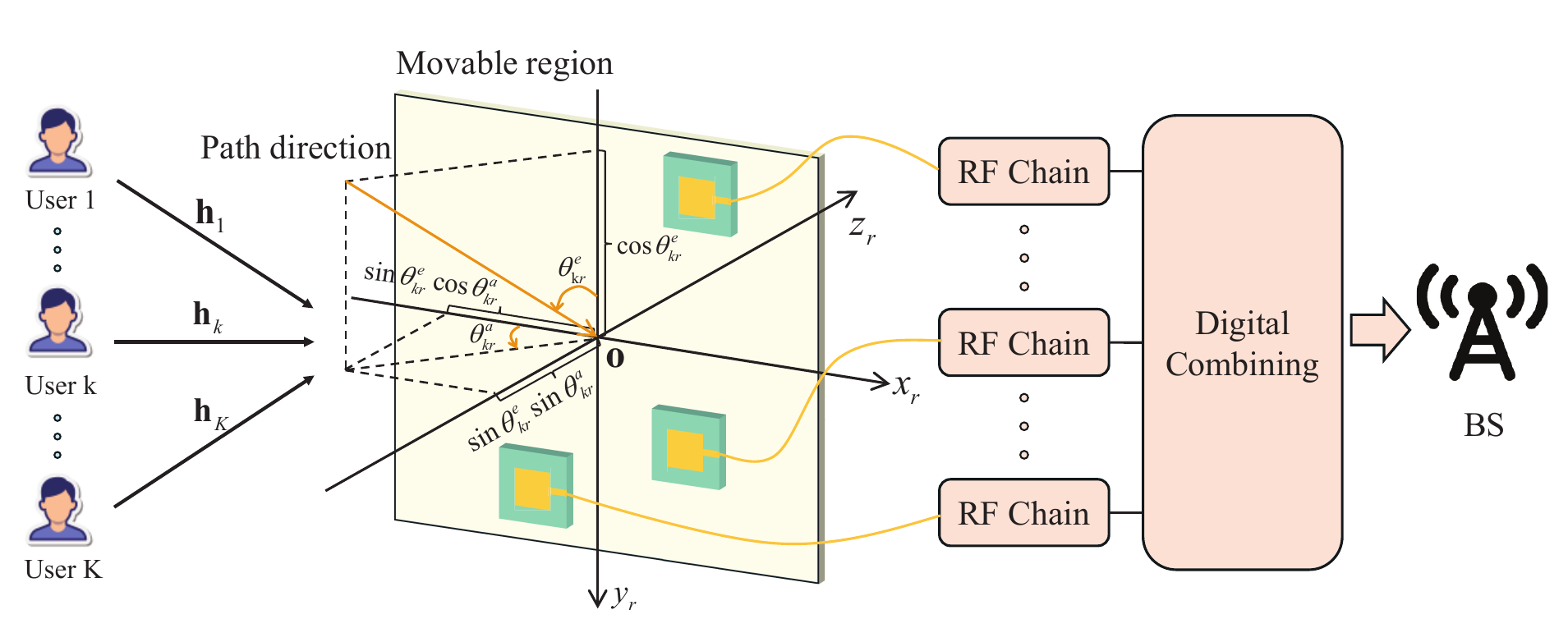}
	\caption{System model: a multiuser MIMO-FAS}
	\label{fig_1}
\end{figure}

As illustrated in Fig. \ref{fig_1}, we consider an uplink transmission for multiuser MIMO-FAS, which consists of the BS equipped with \(M\) fluid antennas and the \(K\) users each equipped with a single transmit antenna.
The positions of fluid antennas can be freely adjusted within a specified two-dimensional (2D) local region \(\mathcal{C}\) since they are connected to radio frequency chains via flexible cables, which allows the precise control of the system through electromechanical drivers such as stepper motors. Without loss of generality, let \(\mathcal{C}\) be a square region of size \(A \times A\).  The positions of all fluid antennas at the BS are represented as  
\(\tilde{\mathbf{t}} = [\mathbf{t}_1, \mathbf{t}_2, \dots, \mathbf{t}_M] \in \mathbb{R}^{2 \times M}\), where 
\(\mathbf{t}_m = [x_m, y_m]^T \in \mathcal{C}\) denotes the \(m\)-th fluid antenna position. The reference point of the region \(\mathcal{C}\) is \(\mathbf{o} = [0,0]^T\). For convenience, we denote the set of all fluid antennas as \(\mathcal{M} = \{1, 2, \dots, M\}\), and the set of all users as 
\(\mathcal{K} = \{1, 2, \dots, K\}\). 
The channel from the user 
\(k, k \in \mathcal{K}\) 
to the BS is denoted by
\(\mathbf{h}_{k} \in \mathbf{C}^{M \times 1}\).
Additionally, we define \(\mathbf{H} = [\mathbf{h}_{1},\mathbf{h}_{2}, \dots, \mathbf{h}_{K}]\).
Therefore, the signal vector received at the BS can be written as
\begin{equation}
	\label{eq:1}
	\mathbf{y} = \sqrt{p}\mathbf{H}\mathbf{x} + \mathbf{n} = \sqrt{p}\sum^{K}_{k=1}{\mathbf{h}_{k} x_{k}} + \mathbf{n}, 
\end{equation}

\noindent where \(p\) denotes the average transmit power of each user, \(\mathbf{x} = [x_{1}, x_{2}, \dots, x_{K}]^T\) represents the information symbol vector of \(K\) users, and \(\mathbf{n} \sim \mathcal{CN}(\mathbf{0}, \sigma^2\mathbf{I}_M)\) is the additive white Gaussian noise (AWGN) with average power \(\sigma^2\).

We utilize a computationally efficient MRC technique to detect the information symbols at the BS. Before constructing the MRC matrix, the BS needs to estimate the channel \(\mathbf{H}\). To this end, a conventional LMMSE estimator is applied, yielding the estimated channel \(\hat{\mathbf{H}}\) based on some predefined pilot signals. We use \(\tau_c\) to represent the length of the channel coherence interval, and \(\tau\) to represent the number of time slots dedicated to channel estimation, where \(\tau\) satisfies \(\tau \geq K\). The details of the channel estimation will be elaborated in the next section. Based on the channel estimation, the BS performs MRC detection by multiplying the received signal \(\mathbf{y}\) with \(\hat{\mathbf{H}}^H\), which is given by
\begin{equation}
	\label{eq:2}
	\mathbf{r} = \hat{\mathbf{H}}^H\mathbf{y} = \sqrt{p} \hat{\mathbf{H}}^H\mathbf{H}\mathbf{x} + \hat{\mathbf{H}}^H\mathbf{n},
\end{equation}

\noindent where \(\mathbf{r}\) represents the resulting signal after processing. Then, the \(k\)-th element of \(\mathbf{r}\) represents the detection signal for the \(k\)-th user, which can be further expressed as 
\begin{equation}
	\label{eq:3}
	\begin{aligned}
		r_{k} &= \sqrt{p} \, \hat{\mathbf{h}}_{k}^{H}\mathbf{h}_{k}x_k + \sqrt{p} \sum_{\substack{i=1 , i\neq k}}^{K} \hat{\mathbf{h}}_{k}^{H}\mathbf{h}_{i}x_i + \hat{\mathbf{h}}_{k}^{H}\mathbf{n} \\
		&= \underbrace{\sqrt{p} \, \mathbb{E}\left\{\hat{\mathbf{h}}_{k}^{H} \mathbf{h}_{k}\right\} x_k}_{\text{Desired signal}} + \underbrace{\sqrt{p}\left(\hat{\mathbf{h}}_{k}^{H} \mathbf{h}_{k} - \mathbb{E}\left\{\hat{\mathbf{h}}_{k}^{H} \mathbf{h}_{k}\right\}\right) x_k}_{\text{Signal leakage}} \\
		& \quad + \underbrace{\sqrt{p} \sum_{\substack{i=1, i\neq k}}^{K} \hat{\mathbf{h}}_{k}^{H} \mathbf{h}_{i} x_i}_{\text{Multi-user interference}} + \underbrace{\hat{\mathbf{h}}_{k}^{H} \mathbf{n}}_{\text{Noise}}, \quad k \in \mathcal{K},
	\end{aligned}
\end{equation}

\noindent where \(\hat{\mathbf{h}}_{k}\) denotes the \(k\)-th column of \(\hat{\mathbf{H}}\).

\subsection{Channel Model}
We adopt the Rician fading model to describe the channel between user \(k\) and the BS, as follows
\begin{equation}
	\label{eq:4}
	\begin{aligned}
		\mathbf{h}_{k} &= \sqrt{\frac{\alpha_k}{\varepsilon_k +1}\varepsilon_k}\bar{\mathbf{h}}_k + \sqrt{\frac{\alpha_k}{\varepsilon_k +1}}\tilde{\mathbf{h}}_k\\
		& = \sqrt{c_k \varepsilon_k}\bar{\mathbf{h}}_k + \sqrt{c_k}\tilde{\mathbf{h}}_k, 
	\end{aligned}
\end{equation}

\noindent where \(\alpha_k\) denotes the path-loss coefficient, \(\varepsilon_k\) denotes the Rician factor, \(c_k\triangleq\frac{\alpha_k}{\varepsilon_k + 1}\) is auxiliary variable, \(\bar{\mathbf{h}}_k \in \mathbb{C}^{M \times 1}\) represents the LoS component, and \(\tilde{\mathbf{h}}_k \in \mathbb{C}^{M \times 1}\) represents the NLoS component. Specifically, for the NLoS path \(\tilde{\mathbf{h}}_k\), the entries are independent and identically distributed (i.i.d) complex Gaussian random variables with zero mean and unit variance.

To better understand the LoS component \(\bar{\mathbf{h}}_k\), we consider the characteristics of this system. It should be noted that this system works at a high frequency band, and the typical size of the antenna moving region is on the order of several to tens of wavelengths. Thus, we assume the size of the fluid antennas' movement region is significantly smaller than the signal propagation distance between the BS and the users, which makes the far-field condition easy to hold. In other words, for each channel path, AoDs, AoAs, and the amplitude of the complex channel coefficient are invariant, while only the phase of each path coefficient changes as a function of the fluid antenna positions.

Then, the difference in signal propagation distance for the LoS path between the \(m\)-th fluid antenna position \(\mathbf{t}_m\) and the reference position \(\mathbf{o}\) is given by 
\begin{equation}
	\label{eq:5}
	\begin{aligned}
		\rho_k^r(\mathbf{t}_m) &\triangleq \mathbf{t}_m^T 
		[\sin\theta_{kr}^e \cos\theta_{kr}^a, \cos\theta_{kr}^e]^T \\
		&= x_m \sin\theta_{kr}^e \cos\theta_{kr}^a + y_m \cos\theta_{kr}^e,
	\end{aligned}
\end{equation}

\noindent where \(\theta_{kr}^e \in [0, \pi]\) and \(\theta_{kr}^a \in [0, \pi]\) are the elevation and azimuth AoAs of the incident signal at the BS from the user \(k\), respectively. Accordingly, the resulting phase difference of the LoS path is \(\frac{2\pi}{\lambda}\rho_k^r(\mathbf{t}_m)\), where \(\lambda\) denotes the carrier wavelength. Hence, for the LoS path, the channel vector 
\(\bar{\mathbf{h}}_k (\mathbf{t})\) between the \(k\)-th user and the BS is given by
\begin{equation}\label{eq:6}
	\bar{\mathbf{h}}_k (\tilde{\mathbf{t}}) = [e^{j\frac{2\pi}{\lambda}\rho_k^r(\mathbf{t}_1)}, e^{j\frac{2\pi}{\lambda}\rho_k^r(\mathbf{t}_2)}, \dots, e^{j\frac{2\pi}{\lambda}\rho_k^r(\mathbf{t}_M)}]^T.
\end{equation}

\subsection{Problem Formulation}
For analytical tractability, we utilize the UatF bound \cite{ref35}, a practical lower bound, to characterize the capacity of the MIMO system. Thus, the lower bound of the \(k\)-th user's ergodic rate can be expressed as
\begin{equation}\label{eq:7}
	\underline{R}_k ( \tilde{\mathbf{t}}) = \frac{\tau_{c} - \tau} {\tau_c} \log_{2} \left( 1 + \text{SINR}_{k} \right),
\end{equation}

\noindent where the pre-log factor \(\frac{\tau_{c} - \tau} {\tau_c}\) represents the rate loss due to the pilot overhead, and the \(\text{SINR}_{k}\) can be expressed as (\ref{eq:8}) at the top of the next page.


\begin{figure*}[ht] 
	\begin{equation}
		\label{eq:8}
		\newcounter{TempEqCnt_1} 
		\setcounter{TempEqCnt_1}{\value{equation}} 
		\setcounter{TempEqCnt_1}{9} 
		\text{SINR}_{k} = 
		\frac{
			p \left| \mathbb{E} \left\{ \hat{\mathbf{h}}_{k}^{H} \mathbf{h}_{k} \right\} \right|^2
		}{
			p \left( \mathbb{E} \left\{ \left| \hat{\mathbf{h}}_{k}^{H} \mathbf{h}_{k} \right|^2 \right\} 
			- \left| \mathbb{E} \left\{ \hat{\mathbf{h}}_{k}^{H} \mathbf{h}_{k} \right\} \right|^2 \right)
			+ p\sum_{\substack{i=1 , i\neq k}}^{K} \mathbb{E} \left\{ \left| \hat{\mathbf{h}}_{k}^{H} \mathbf{h}_{i} \right|^2 \right\} 
			+ \sigma^2 \mathbb{E} \left\{ \left\| \hat{\mathbf{h}}_{k} \right\|^2 \right\}
		}.
	\end{equation}
	\hrulefill  
	\newcounter{TempEqCnt_2} 
	\setcounter{TempEqCnt_2}{\value{equation}} 
	\setcounter{equation}{19} 
	\begin{align}
		&E_k^{signal} = \left\{E_k^{noise}\right\}^2, \label{eq:20}\\
		&E_k^{noise} = Mc_k\left(\varepsilon_k + a_k\right), \label{eq:21}\\
		&E_k^{leak} = M^2c_k^2\varepsilon_k^2 + Mc_k^2\varepsilon_k + Ma_k^2c_k^2\varepsilon_k + a_k^2c_k^2M(M+1) + \frac{\sigma^2}{\tau p}Ma_k^2c_k\varepsilon_k + \frac{\sigma^2}{\tau p}a_k^2c_k + 2M^2c_k^2a_k\varepsilon_k - M^2c_k^2(\varepsilon_k+a_k)^2, 
		\label{eq:22}\\
		&I_{ki} =c_kc_i\varepsilon_k\varepsilon_i\left|f_k\left(\mathbf{t}\right)\right|^2 
		+ Mc_kc_i\varepsilon_k 
		+ Ma_k^2c_kc_i\varepsilon_i 
		+ Ma_k^2c_kc_i 
		+ \frac{\sigma^2}{\tau p}Ma_k^2c_i\varepsilon_i 
		+ \frac{\sigma^2}{\tau p}a_k^2c_i, 
		\label{eq:23}
	\end{align}
	\hrulefill  
	\setcounter{equation}{\value{TempEqCnt_2}} 
\end{figure*}

In this paper, we aim to maximize the minimum rate of all users through optimizing the matrix of fluid antenna positions \(\tilde{\mathbf{t}}\) in the large timescale, thereby ensuring fairness among different users. The general optimization problem is then formulated as follows
\begin{subequations} \label{eq:9}
	\begin{align}
		\quad & \max_{\tilde{\mathbf{t}}} \min_{k \in \mathcal{K}} \quad \underline{R}_k ( \tilde{\mathbf{t}}), \label{eq:9a} \\
		& \: \text{s.t.} \quad \|\mathbf{t}_m - \mathbf{t}_i\| \geq D_\text{min}, \ \forall m, i \in \mathcal{M}, m \neq i, \label{eq:9b} \\
		&  \quad \quad \quad \mathbf{t}_m \in \mathcal{C}, \ \forall m \in \mathcal{M}, \label{eq:9c}
	\end{align}
\end{subequations}

\noindent where constraint (\ref{eq:9b}) ensures the minimum spacing distance \(D_\text{min}\) between any two adjacent fluid antennas to avoid coupling effects, and constraint (\ref{eq:9c}) indicates the feasible region of fluid antenna position. Notably, solving this max-min problem is highly challenging, because it involves not only a non-convex objective function (\ref{eq:9a}) and constraint (\ref{eq:9c}), but also the absence of a closed-form expression for the ergodic rate.

\section{Channel Estimation} \label{sec3}
The BS estimates the CSI via the LMMSE method, utilizing received pilot signals occupying $\tau$ symbols within the channel coherence interval for uplink training. During the training process, all users simultaneously transmit mutually orthogonal pilot sequences, satisfying the orthogonality condition $\mathbf{S}^H\mathbf{S} = \mathbf{I}_K$ where $\mathbf{S} = \left[ \mathbf{s}_1, \mathbf{s}_2, \dots, \mathbf{s}_K\right] \in \mathbb{C}^{\tau \times K}$ denotes the pilot matrix. Each transmitted pilot symbol carries power $p$, yielding total pilot power $\tau p$ per user. Then, the $M\times \tau$ pilot signal matrix received at the BS is expressed as 

\begin{equation}\label{eq:10}
	\mathbf{Y}_p = \sqrt{\tau p}\mathbf{H}\mathbf{S}^H + \mathbf{N} = \sqrt{\tau p} \sum_{i=1}^{K}\mathbf{h}_i\mathbf{s}_i^H + \mathbf{N}, 
\end{equation}

\noindent where $\mathbf{N} \in \mathbb{C}^{M\times\tau}$ is a noise matrix with i.i.d $\mathcal{CN}\left(0, \sigma^2\right)$ elements. The observation signal for user $k$ is obtained through multiplying (\ref{eq:10}) by $\frac{\mathbf{s}_k}{\sqrt{\tau p}}$, as follows
\begin{equation}
	\label{eq:11}
	\mathbf{y}_p^k = \frac{1}{\sqrt{\tau p}}\mathbf{Y}_p\mathbf{s}_k=\mathbf{h}_k + \frac{\mathbf{N} \mathbf{s}_k}{\sqrt{\tau p}}.
\end{equation}

\begin{theorem} 
	\label{theo-1}
	For $k\in\mathcal{K}$, the estimated channel vector $\hat{\mathbf{h}}_k$ obtained from the observation signal $\mathbf{y}_p^k$ based on LMMSE estimator can be written as
	\begin{align}
		\hat{\mathbf{h}}_k &= a_k \mathbf{y}_p^k + \mathbf{b}_k \nonumber \\
		&= \sqrt{c_k \varepsilon_k} \bar{\mathbf{h}}_k + a_k \sqrt{c_k} \tilde{\mathbf{h}}_k + \frac{a_k}{\sqrt{\tau p}} \mathbf{N} \mathbf{s}_k, \label{eq:12} \\
		a_k &= \frac{c_k}{c_k + \frac{\sigma^2}{\tau p}}, \label{eq:13} \\
		\mathbf{b}_k &= \left(1 - a_k\right) \sqrt{c_k \varepsilon_k} \bar{\mathbf{h}}_k. \label{eq:14}
	\end{align}
\end{theorem}

\itshape {Proof:}  \upshape See Apendix \ref{appendix_A}.

\section{Analysis of the achievable rate} \label{sec4}
In this section, the closed-form expression for a lower bound of the achievable rate is obtained by combining (\ref{eq:7}) with derived specific $\mathrm{SINR}_k$ expression, which exploits the estimated channel provided in Theorem \ref{theo-1} and the statistical CSI. The expression of $\mathrm{SINR}_k$ is given in the following theorem.
\begin{theorem} 
	\label{theo-2}
	The $\mathrm{SINR}$ for user $k,k\in\mathcal{K}$ in (\ref{eq:7}) is given by
	\begin{equation}
		\label{eq:15}
		\mathrm{SINR}_k =  \frac{pE_k^{signal}}{pE_k^{leak}+p\sum_{\substack{i=1 , i\neq k}}^{K}I_{ki}+\sigma^2E_k^{noise}} 
	\end{equation}
	where
	\begin{align}
		&E_k^{signal} \triangleq \left|\mathbb{E}\left\{\hat{\mathbf{h}}_k^H \mathbf{h}_k\right\}\right|^2, 
		\label{eq:16} \\
		&E_k^{leak} \triangleq \mathbb{E}\left\{\left|\hat{\mathbf{h}}_k^H\mathbf{h}_k\right|^2\right\} - \left|\mathbb{E}\left\{\hat{\mathbf{h}}_k^H \mathbf{h}_k \right\}\right|^2, 
		\label{eq:17} \\
		&I_{ki} \triangleq \mathbb{E}\left\{\left|\hat{\mathbf{h}}_k^H \mathbf{h}_i\right|^2\right\}, 
		\label{eq:18} \\
		&E_k^{noise} \triangleq \mathbb{E}\left\{\left\|\hat{\mathbf{h}}_k\right\|^2\right\}, 
		\label{eq:19}
	\end{align}
	are respectively given in (\ref{eq:20}) - (\ref{eq:23}), shown at the top of the previous page, with 
	\begin{align}
		\label{eq:24}
		\setcounter{equation}{23} 
		\left|f_k(\tilde{\mathbf{t}})\right|^2= \left|\bar{\mathbf{h}}_k^H\bar{\mathbf{h}}_i\right|^2.
	\end{align}
\end{theorem}

\itshape {Proof:}  \upshape See Apendix \ref{appendix_B}.


\section{Optimization of the antenna position} \label{sec5}
Optimizing the positions of fluid antennas presents significant challenges. On the one hand, the strong nonlinearity of $\underline{R}_k ( \tilde{\mathbf{t}})$, coupled with the non-differentiable property of the max-min fairness objective, renders conventional optimization methods such as semidefinite relaxation (SDR) and majorization-minimization (MM) ineffective. On the other hand, the large solution space of the $\tilde{\mathbf{t}}$. i.e., $[-A/2, A/2]^{2M}$, makes exhaustive search methods prohibitive computational complexity. To address these challenges, we first propose a GA-based method. Although it is well-suited for the non-convex problem and search spaces, effectively avoiding local optima, the GA suffers from high computational complexity and slow convergence speed. Additionally, we propose an accelerated gradient ascent algorithm, which offers lower complexity and faster convergence when the objective function is differentiable, although it remains susceptible to local optima.

\subsection{GA-based Solution}

Rooted in biological evolutionary principles, GA emulates the evolution of a population in nature. In the GA-based approach, we first initialize N individuals forming the initial population $\mathcal{T}^{(0)} = \left\{\tilde{\mathbf{t}}^{(0)}_1, \tilde{\mathbf{t}}^{(0)}_2, ..., \tilde{\mathbf{t}}^{(0)}_N\right\}$, where each individual represents potential antenna coordinates as follows
\begin{align}
	&\tilde{\mathbf{t}}^{(0)}_n = [\mathbf{t}^{(0)}_{n,1}, \mathbf{t}^{(0)}_{n,2}, ..., \mathbf{t}^{(0)}_{n,M}], 
	\label{eq:25}\\
	&\mathbf{t}^{(0)}_{n,m} = [x^{(0)}_{n,m}, y^{(0)}_{n,m}]^T,
	\label{eq:26}
\end{align}

\noindent where $x^{(0)}_{n,m}, y^{(0)}_{n,m} \sim \mathcal{U}[-A/2,A/2]$ for $ 1 \leq n \leq N, 1 \leq m \leq M$ guarantee that the constraint (\ref{eq:9c}) holds. Additionally, to ensure that at least one individual has a positive fitness value, the position matrices of some individuals are initialized based on the uniform planar array (UPA) geometry with guaranteed minimum inter-antenna distance $D_{min}$.

The fitness function incorporates adaptive penalty term to enforce constraint (\ref{eq:9c}), which is given by
\begin{equation}
	\label{eq:27}
	\mathcal{F}\left\{\tilde{\mathbf{t}}^{(i)}_n\right\} = \underline{R}_k \left(\tilde{\mathbf{t}}^{(i)}_n\right) - \omega\left|\mathcal{V}\left(\tilde{\mathbf{t}}^{(i)}_n\right)\right|,
\end{equation}

\noindent where $i$ denotes iteration index, $\omega$ is a large positive penalty parameter, and $\mathcal{V}\left(\tilde{\mathbf{t}}^{(i)}_n\right)$ is a defined set, whose elements correspond to a pair positions of fluid antennas in $\tilde{\mathbf{t}}$ violating the minimum spacing distance $ D_{min}$, as follows
\begin{align}
	\label{eq:28}
	&\mathcal{V}\left(\tilde{\mathbf{t}}\right) \nonumber\\
	&= \left\{ \left( \mathbf{t}_m, \mathbf{t}_{m'} \right)| \left\|\mathbf{t}_m - \mathbf{t}_{m'}\right\|_2 < D_{min}, 1 \leq m < m' \leq M \right\}.
\end{align}

The penalty parameter $\omega$ needs to ensure that $\underline{R}_k \left(\tilde{\mathbf{t}}^{(i)}_n\right) - \omega \leq 0$ holds for all individuals. If any individual violates the minimum spacing distance constraint, its fitness value will be less than zero. Therefore, in each iteration, all individuals can move toward positions that satisfy constraint (\ref{eq:9b}).

\subsection{Low-complexity Gradient-based Solution}
Although the GA-based method can simply solve this complicated problem, its computational complexity is relatively high. In this subsection, we adopt Nesterov's accelerated gradient ascent algorithm to solve this problem with lower complexity, achieving comparable performance. It is noted that this algorithm can effectively speed up the convergence rate of the gradient-based method to obtain better performance \cite{ref37}.

Due to the non-differentiable minimum function in the objective function (\ref{eq:9a}), the Lagrangian duality and the Jaynes maximum entropy principle are adopted. Then, an approximation of the minimum achievable user rate can be represented as follows 
\begin{equation}
	\label{eq:29}
	\min_{k \in \mathcal{K}}\underline{R}_k (\tilde{\mathbf{t}}) 
	\approx -\frac{1}{\mu} \text{ln}\left\{\sum_{k=1}^{K}\text{exp}\{-\mu\underline{R}_k (\tilde{\mathbf{t}})\}\right\} 
	\triangleq g(\tilde{\mathbf{t}}),
\end{equation}

\noindent where $\mu$ is a constant value that determines the approximation error to be less than $\frac{\text{ln}K}{\mu}$, which has been proved in \cite{ref36}. Thus, the problem can be recast as 
\begin{align}
	& \max_{\tilde{\mathbf{t}}} \, g(\tilde{\mathbf{t}}) , 
	\label{eq:30} \\
	& \: \text{s.t.} \quad (\ref{eq:9b}), (\ref{eq:9c}). \nonumber
\end{align}

Then, we need to calculate the gradient of $g(\tilde{\mathbf{t}})$. Based on the chain rule, we have
\begin{equation}
	\label{eq:31}
	\frac{\partial g(\tilde{\mathbf{t}})}{\partial \tilde{\mathbf{t}}} = \frac{\tau_c - \tau}{\tau_c} \frac{\sum_{k=1}^{K}\left\{\frac{\text{exp}\left\{-\mu \underline{R}_k (\tilde{\mathbf{t}})\right\}}{1 + \mathrm{SINR}_k(\tilde{\mathbf{t}})}\frac{\partial \mathrm{SINR}_k(\tilde{\mathbf{t}})}{\partial \tilde{\mathbf{t}}}\right\}}{\text{ln}2\left(\sum_{k=1}^{K}\text{exp}\left\{-\mu \underline{R}_k (\tilde{\mathbf{t}}) \right\}\right)},
\end{equation}

\noindent and 
\begin{equation}
	\label{eq:32}
	\frac{\partial \mathrm{SINR}_k(\tilde{\mathbf{t}})}{\partial \tilde{\mathbf{t}}} = -\frac{p^2 E_k^{signal} \sum\limits_{\substack{i=1, i \neq k}}^{K}\frac{\partial I_{ki}(\tilde{\mathbf{t}})}{\partial \tilde{\mathbf{t}}}}{\left(pE_k^{leak}+p\sum\limits_{\substack{i=1, i \neq k}}^{K}I_{ki}(\tilde{\mathbf{t}}) + E_k^{noise}\right)^2}. \end{equation}

Thus, the gradient of $g(\tilde{\mathbf{t}})$ can be obtained after calculating $\frac{\partial I_{ki}(\tilde{\mathbf{t}})}{\partial \tilde{\mathbf{t}}}$.

\begin{lemma} 
	\label{lem-1}
	The gradient of $I_{ki}(\tilde{\mathbf{t}})$ is given by
	\begin{align}
		& \frac{\partial I_{ki}(\tilde{\mathbf{t}})}{\partial \tilde{\mathbf{t}}} = 2c_k c_i \epsilon_k\epsilon_i\mathrm{Re}\left\{\frac{\partial f_k(\tilde{\mathbf{t}})}{\partial \tilde{\mathbf{t}}}f_k^*(\tilde{\mathbf{t}})\right\},
		\label{eq:33} \\
		&\left[\frac{\partial f_k(\tilde{\mathbf{t}})}{\partial \tilde{\mathbf{t}}}\right]_{1,u} \nonumber\\ 
		&= j\frac{2\pi}{\lambda}\left(\sin\theta_{ir}^e \cos\theta_{ir}^a - \sin\theta_{kr}^e \cos\theta_{kr}^a\right)e^{j\frac{2\pi}{\lambda}\left[ \rho_i^r(\mathbf{t}_u) - \rho_k^r(\mathbf{t}_u) \right]}, \nonumber\\
		&\quad\quad\quad\quad\quad\quad\quad\quad\quad\quad\quad\quad\quad\quad\quad\quad 1\leqslant u\leqslant M,
		\label{eq:34} \\
		&\left[\frac{\partial f_k(\tilde{\mathbf{t}})}{\partial \tilde{\mathbf{t}}}\right]_{2,u} \nonumber \\ 
		&= j\frac{2\pi}{\lambda}\left(\cos\theta_{ir}^e - \cos\theta_{kr}^e \right) e^{j\frac{2\pi}{\lambda}\left[ \rho_i^r(\mathbf{t}_u) - \rho_k^r(\mathbf{t}_u) \right]}, \nonumber\\
		&\quad\quad\quad\quad\quad\quad\quad\quad\quad\quad\quad\quad\quad\quad\quad\quad 1\leqslant u\leqslant M.
		\label{eq:35}
	\end{align}
\end{lemma}

\itshape {Proof:}  \upshape First, we consider the gradient of $I_{ki}(\tilde{\mathbf{t}})$ through the chain rule.
\begin{align}
	\label{eq:36}
	\frac{\partial I_{ki}(\tilde{\mathbf{t}})}{\partial \tilde{\mathbf{t}}} 
	&=c_kc_i\varepsilon_k\varepsilon_i \frac{\partial}{\partial \tilde{\mathbf{t}}}\left(\left|f_k\left(\tilde{\mathbf{t}}\right)\right|^2\right) \nonumber\\
	&= c_k c_i \epsilon_k\epsilon_i \frac{\partial}{\partial \tilde{\mathbf{t}}}\left(f_k(\tilde{\mathbf{t}})f_k^*(\tilde{\mathbf{t}})\right) \nonumber\\
	&=c_k c_i \epsilon_k\epsilon_i\left( \frac{\partial f_k(\tilde{\mathbf{t}})}{\partial \tilde{\mathbf{t}}}f_k^*(\tilde{\mathbf{t}}) + f_k(\tilde{\mathbf{t}})\frac{\partial f_k^*(\tilde{\mathbf{t}})}{\partial \tilde{\mathbf{t}}} \right) \nonumber\\
	&=2c_k c_i \epsilon_k\epsilon_i\mathrm{Re}\left\{\frac{\partial f_k(\tilde{\mathbf{t}})}{\partial \tilde{\mathbf{t}}}f_k^*(\tilde{\mathbf{t}})\right\},
\end{align}
where $f_k^*(\tilde{\mathbf{t}})$ denotes the conjugate of $f_k(\tilde{\mathbf{t}})$ .Therefore, (\ref{eq:33}) is proved.

Then, $\left[\frac{\partial f_k(\tilde{\mathbf{t}})}{\partial \tilde{\mathbf{t}}}\right]_{1,u}, \left[\frac{\partial f_k(\tilde{\mathbf{t}})}{\partial \tilde{\mathbf{t}}}\right]_{2,u}, 1\leqslant u\leqslant M$ represent the elements of the first row and the second row of the gradient matrix of $f_k(\tilde{\mathbf{t}})$, respectively, which are given by
\begin{align}
	&\left[\frac{\partial f_k(\tilde{\mathbf{t}})}{\partial \tilde{\mathbf{t}}}\right]_{1,u} 
	= \left[\frac{\partial \bar{\mathbf{h}}_k^H(\tilde{\mathbf{t}})\bar{\mathbf{h}}_i(\tilde{\mathbf{t}})}{\partial \tilde{\mathbf{t}}}\right]_{1,u} 
	= \frac{\partial \bar{\mathbf{h}}_k^H(\tilde{\mathbf{t}})\bar{\mathbf{h}}_i(\tilde{\mathbf{t}})}{\partial x_u}\nonumber\\ 
	&= \frac{\partial }{\partial x_u} \sum_{m=1}^{M}e^{j\frac{2\pi}{\lambda}\left[ \rho_i^r(\mathbf{t}_m) - \rho_k^r(\mathbf{t}_m) \right]} \nonumber\\ 
	&= j\frac{2\pi}{\lambda}\left(\sin\theta_{ir}^e \cos\theta_{ir}^a - \sin\theta_{kr}^e \cos\theta_{kr}^a\right)e^{j\frac{2\pi}{\lambda}\left[ \rho_i^r(\mathbf{t}_u) - \rho_k^r(\mathbf{t}_u) \right]},
	\label{eq:37} \\
	&\left[\frac{\partial f_k(\tilde{\mathbf{t}})}{\partial \tilde{\mathbf{t}}}\right]_{2,u} 
	= \left[\frac{\partial \bar{\mathbf{h}}_k^H(\tilde{\mathbf{t}})\bar{\mathbf{h}}_i(\tilde{\mathbf{t}})}{\partial \tilde{\mathbf{t}}}\right]_{2,u} 
	= \frac{\partial \bar{\mathbf{h}}_k^H(\tilde{\mathbf{t}})\bar{\mathbf{h}}_i(\tilde{\mathbf{t}})}{\partial y_u} \nonumber \\ 
	&= \frac{\partial }{\partial y_u} \sum_{m=1}^{M}e^{j\frac{2\pi}{\lambda}\left[ \rho_i^r(\mathbf{t}_m) - \rho_k^r(\mathbf{t}_m) \right]} \nonumber\\ 
	&= j\frac{2\pi}{\lambda}\left(\cos\theta_{ir}^e - \cos\theta_{kr}^e \right) e^{j\frac{2\pi}{\lambda}\left[ \rho_i^r(\mathbf{t}_u) - \rho_k^r(\mathbf{t}_u) \right]}.
	\label{eq:38}
\end{align}
Thus, (\ref{eq:34}) and (\ref{eq:35}) are established, which completes the proof.

By substituting (\ref{eq:33}), (\ref{eq:34}), and (\ref{eq:35}) into (\ref{eq:31}) and (\ref{eq:32}), the final analytical expression for the gradient of $ g\left(\tilde{\mathbf{t}}\right) $ can be obtained.

To guarantee the constraint in (\ref{eq:9c}), each update for the matrix of antenna positions is followed by a projection operation. Define $\mathcal{Y}\left\{ \tilde{\mathbf{t}}\right\}$ as projection function as follows 
\begin{align}
	\label{eq:39}
	\left[\mathcal{Y}(\tilde{\mathbf{t}})\right]_{u,m}=
	\begin{cases}
		-\frac{A}{2},&\text{if} \  [\tilde{\mathbf{t}}]_{u,m}<-\frac{A}{2}, \\
		\frac{A}{2}, &\text{if} \  [\tilde{\mathbf{t}}]_{u,m}>\frac{A}{2}, \\
		[\tilde{\mathbf{t}}]_{u,m},&\text{otherwise}, 
	\end{cases}
\end{align}
for $1 \leq u \leq 2, 1 \leq m \leq M$.

\begin{algorithm}[t]
	\renewcommand{\algorithmicrequire}{\textbf{Input:}}
	\renewcommand{\algorithmicensure}{\textbf{Output:}}
	\caption{Accelerated Gradient Ascent Algorithm}
	\label{algorithm1}
	\begin{algorithmic}[1]
		\STATE Initialize antenna positions $\tilde{\mathbf{t}}^{(0)}$ randomly and set $i=0,l^{(0)}=0.5,\mathbf{v}^{(-1)}=\tilde{\mathbf{t}}^{(0)}$;
		\WHILE {1} 
		\STATE Calculate the gradient matrix $\nabla _{\tilde{\mathbf{t}}} g\left( \tilde{\mathbf{t}}^{(i)} \right)$;
		\STATE Obtain the step size $\zeta^{(i)}$ based on the backtracking line search until satisfying constraints (\ref{eq:40a}) and (\ref{eq:40b});
		
		\STATE $\mathbf{v}^{(i)}=\mathcal{Y}
		\left\{\tilde{\mathbf{t}}^{(i)}+\zeta^{(i)} \nabla _{\tilde{\mathbf{t}}} g\left(\tilde{\mathbf{t}}^{(i)}\right)\right\}$;
		\STATE $l^{(i+1)}=(1+\sqrt{4{l^{(i)}}^2+1})/2$;
		\STATE $\tilde{\mathbf{t}}^{(i+1)}=\mathcal{Y}
		\left\{\mathbf{v}^{(i)}+(l^{(i)}-1)(\mathbf{v}^{(i)}-\mathbf{v}^{(i-1)})/l^{(i+1)}\right\}$;
		
		\IF {$g(\tilde{\mathbf{t}}^{(i+1)})-g(\tilde{\mathbf{t}}^{(i)})<10^{-4}$}
		\STATE $\tilde{\mathbf{t}}^*=\tilde{\mathbf{t}}^{(i+1)}$, break;
		\ENDIF
		
		\STATE $i=1+1$;
		
		\ENDWHILE	
	\end{algorithmic}
\end{algorithm}

We assume the position matrix in the $(i)$-th iteration as $\tilde{\mathbf{t}}^{(i)}$, and the gradient of function $g\left(\tilde{\mathbf{t}}| \tilde{\mathbf{t}}^{(i)}\right)$ at the antenna position matrix $\tilde{\mathbf{t}}^{(i)}$ in the $(i)$-th iteration as $\nabla _{\tilde{\mathbf{t}}} g\left(\tilde{\mathbf{t}}^{(i)}\right)= \frac{\partial g(\tilde{\mathbf{t}})}{\partial \tilde{\mathbf{t}}}|_{\tilde{\mathbf{t}} = \tilde{\mathbf{t}}^{(i)}}$. 
The convergence behavior of the gradient ascent method highly depends on the step length, which can be calculated based on the backtracking line search. To be specific, the initial step length is set a fixed value, and it iteratively decreases during the search process with a scaling factor $\kappa\in(0,1),\mathrm{~i.e.,~}\zeta^{(i)}\leftarrow\kappa\zeta^{(i)}$, until satisfies the Armijo-Goldstein condition and the minimum spacing constraint, as follows

\begin{subequations}
	\begin{align}
		& g_{i-1}(\tilde{\mathbf{t}}^{(i)})\geq g_{i-1}(\tilde{\mathbf{t}}^{(i-1)})+\varpi\zeta^{(i)}\left\|\nabla_{\tilde{\mathbf{t}}}g_{i-1}\left(\tilde{\mathbf{t}}^{(i-1)}\right)\right\|^{2}, 
		& &  \label{eq:40a}\\
		& \left|\mathcal{P}\left(\tilde{\mathbf{t}}^{(i)}\right)\right|=0, & \label{eq:40b}
	\end{align}
\end{subequations}
 where $\varpi \in (0,1)$ denotes a control parameter to evaluate whether the step size is appropriate to enable the function $g(\tilde{\mathbf{t}}^{(i)})$ to increase efficiently. 
 The main step of the low-complexity projected accelerated gradient ascent method is provided in Algorithm 1, where lines 6 and 7 correspond to the accelerated process.

\section{Numerical Results} \label{sec6}
In this section, numerical results of the multiuser MIMO-FAS with imperfect CSI are provided to validate the effectiveness of our two algorithms. We also analyze the impact of key parameters. 
The simulation considers a BS equipped with $M = 9$ fluid antennas serving $K = 3, 5, 7$ users, which are randomly distributed within $d_k \sim \mathcal{U}[50, 70]$ meters. The elevation and azimuth angles of AoDs and AoAs for each user and the BS are evenly distributed in the range of [0, $\lambda$]. Additionally, the movable region at the BS is set as $\mathcal{C} = [-A/2, A/2] \times [-A/2, A/2]$, and the minimum spacing between two adjacent fluid antennas is set as $D_{\text{min}} = \frac{\lambda}{2}$. We set the large-scale path-loss coefficients equal to $\alpha_k = \alpha_0 d_k^{-2.8}, \forall k$, where $\alpha_0 = -40$ dB indicates the reference average path loss at $1$ m. The Rician fading channel is modeled with a consistent Rician factor $\varepsilon_k = \varepsilon = 6, \forall k$, modeling moderate LoS condition. We assume that the length of the channel coherence interval as $\tau_{c} = 196$, and  the number of symbols for channel estimation as $\tau = K$. The noise power is $\sigma^2 = -104$ dBm. Table \ref{tab:table1} displays other simulation parameters and further specific parameters of the simulation are  mentioned in the following.
\begin{table}[!t]
	\caption{Simulation parameters\label{tab:table1}}
	\renewcommand{\arraystretch}{1.5} 
	\centering
	\begin{tabular}{|p{1.5cm}|p{4cm}|p{1.5cm}|}
		\hline
		\textbf{Parameter} & \textbf{Description} &\textbf{Value}\\
		\hline
		$\lambda$ & Carrier wavelength & 0.1 m\\
		\hline
		$A$ & Size of movable region & 6$\lambda$\\
		\hline
		$p$ & Transmit power & 30 dBm\\
		\hline
		$\mu$ & Approximation factor & 100\\
		\hline
		$\omega$ & Penalty factor & 10\\
		\hline
		$\kappa$ & Scaling factor & 0.8\\
		\hline
		$\varpi$ & Control parameter & 0.5\\
		\hline
	\end{tabular}
\end{table}

Two proposed approaches are labeled by “$\textbf{FAS-GA}$” and “$\textbf{FAS-gradient}$”, which we compare with the following benchmark.
\begin{itemize}
	\item{$\textbf{FPA}$}: The antennas at the BS are uniformly planar arrays with inter-element antenna spacing $\frac{\lambda}{2}$.
\end{itemize}

\begin{figure}[!t]
	\centering
	\includegraphics[width=3.2in]{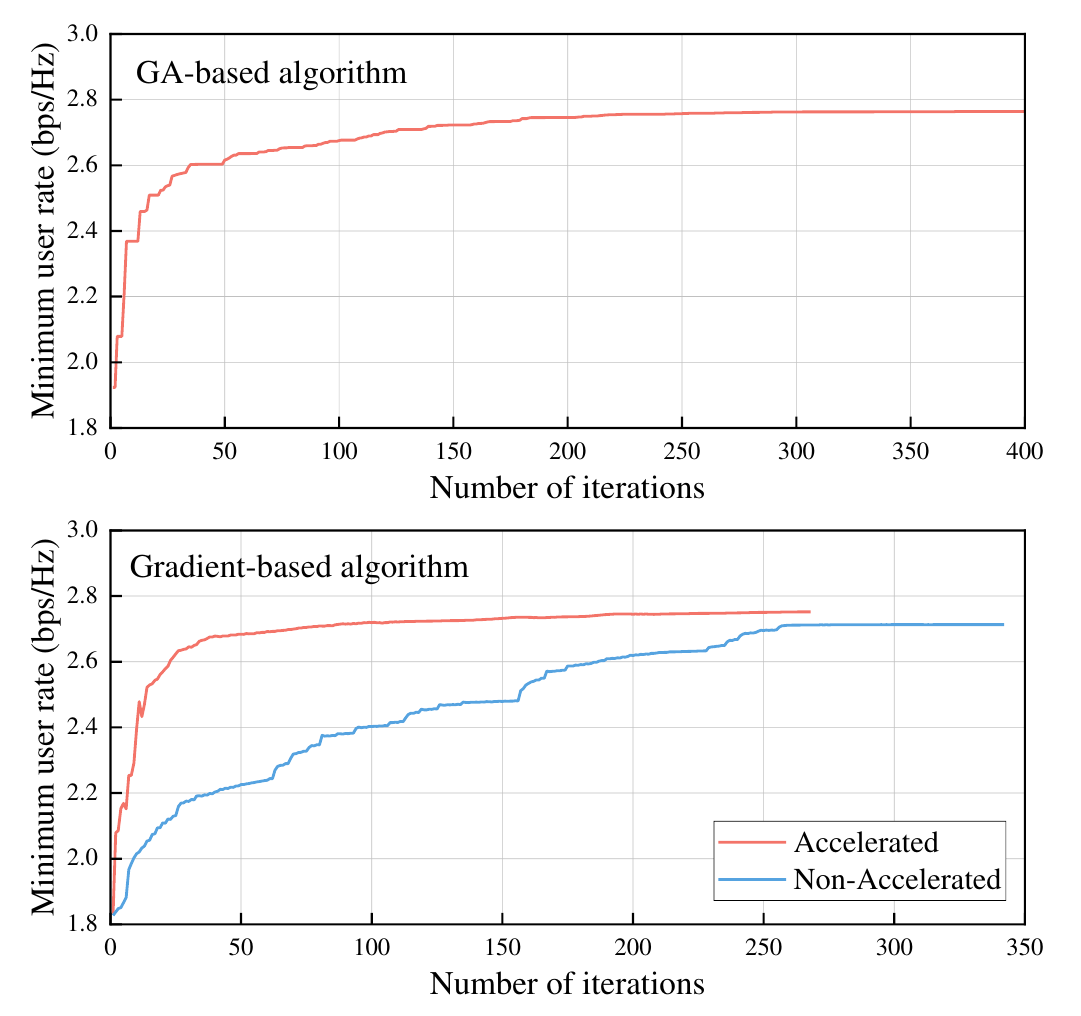}
	\caption{Convergence performance of GA-based and Gradient-based algorithms.}
	\label{fig_2}
\end{figure}

Fig. \ref{fig_2} illustrates the convergence behavior of the proposed FAS-GA scheme and the FAS-gradient scheme in a scenario with $M = 9$ fluid antennas at the BS and  $K = 5$ users, and FAS-gradient scheme is further compared with the non-accelerated counterpart.
As can be observed, the accelerated gradient approach converges within 100 times. Compared with the at least 250 iterations that use the non-accelerated gradient method, there is a significant enhancement in convergence speed. Additionally, although FAS-GA achieves convergence within 200 iterations, which has not much difference in the number of iterations with FAS-gradient, the total time consumption is reduced and the efficiency increases due to lower computational complexity in each iteration.

\begin{figure}[!t]
	\centering
	\includegraphics[width=\columnwidth]{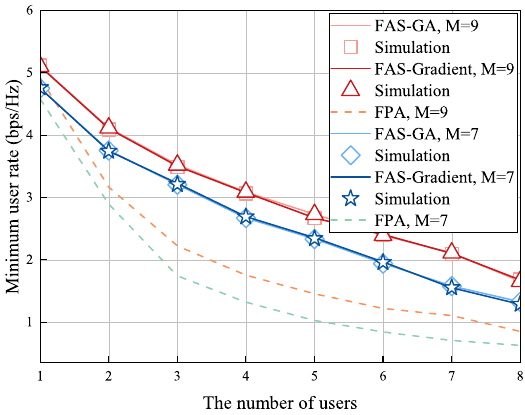}
	\caption{Minimum user rate versus the number of users $K$.}
	\label{fig_3}
\end{figure}

In Fig. \ref{fig_3}, we explore the relationship between the number of users and achievable rate when $M = 7, 9$, accompanied by Monte Carlo simulation results.
The Monte Carlo simulations are conducted to validate the correctness of mathematical derivations, which are obtained from (\ref{eq:8}) through averaging $10^5$ random channel realizations, denoted as `Simulation' in the figure legends. 
There is an excellent agreement between the approximate achievable rate expression in (\ref{eq:7}) and the simulation results. It is noted that FAS-GA and FAS-gradient provide nearly identical performance, delivering significant rate improvements over conventional FPA systems. 
Moreover, it can be observed that as the number of users $K$ increases, the minimum user rate decreases, indicating that spatial diversity gains from fluid antennas cannot fully counteract intensified multiuser interference. Regardless of the number of users, the 9-antenna configuration consistently outperforms its 7-antenna counterpart.

\begin{figure}[!t]
	\centering
	\includegraphics[width=\columnwidth]{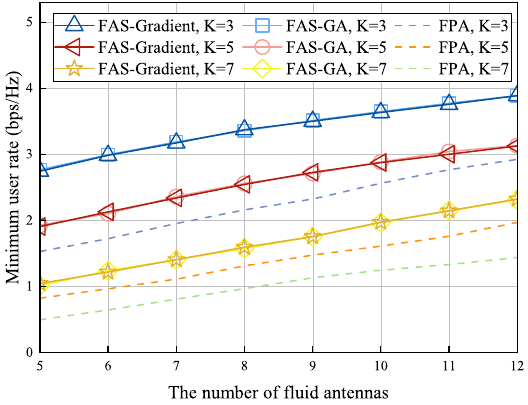}
	\caption{Minimum user rate versus the number of antennas $M$.}
	\label{fig_4}
\end{figure}

For a clearer investigation into the number of fluid antennas, Fig. \ref{fig_4} shows the minimum user rate versus the number of fluid antennas for different user densities $K = 3, 5, 7$. There is an indistinguishable difference in performance between FAS-GA and FAS-gradient, with both consistently outperforming the FPA system. As shown in the figure, the minimum user rate exhibits monotonic growth with the number of fluid antennas across all user configurations. This is attributed to enhanced spatial diversity gain enabled by additional antennas and interference suppression achieved through fluid antenna position optimization that impairs the multiuser interference term (\ref{eq:23}).

\begin{figure}[!t]
	\centering
	\includegraphics[width=\columnwidth]{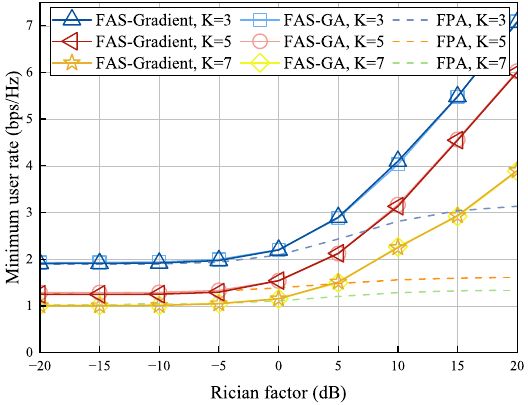}
	\caption{Minimum user rate versus Rician factor $\varepsilon$.}
	\label{fig_5}
\end{figure}

Fig. \ref{fig_5} reveals the influence of the Rician factor when the number of users is different. Across all cases, fluid antenna-enhanced schemes dramatically benefit from the increasing Rician factor $\varepsilon$. In weak LoS channel conditions where $\varepsilon < -5 \text{dB}$, the minimum user rate for FAS-GA, FAS-gradient, and FPA schemes remains unchanged as $\varepsilon$ increases, with the performance of three approaches nearly identical. As the channel becomes more predictable, the fluid antenna-assisted scheme demonstrates a greater optimization potential. When the Rician factor increases from 5 dB to 20 dB, the minimum user rate improves for various user scenarios, increasing from 1.9 bps/Hz to 7.1 bps/Hz for 3 users, from 1.2 bps/Hz to 6.0 bps/Hz for 5 users, and from 1.0 bps/Hz to 3.9 bps/Hz for 7 users.

\begin{figure}[!t]
	\centering
	\includegraphics{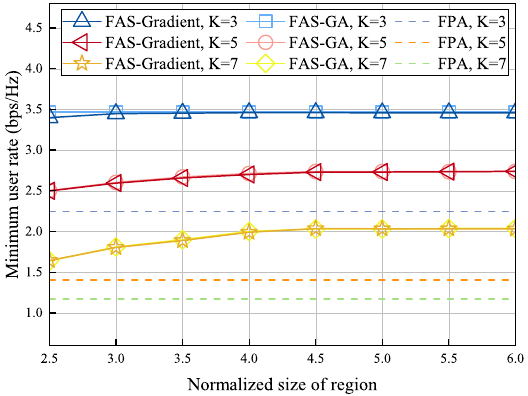}
	\caption{Minimum user rate versus normalized size of region $A/\lambda$.}
	\label{fig_6}
\end{figure}

Finally, Fig. \ref{fig_6} displays the minimum user rate of three schemes for normalized movable region size $A/\lambda$. The performance of FAS-GA and FAS-gradient is closely matched. For $K = 7$, as the movable region increases, the minimum user rate of the fluid antenna-enhanced system rise from 1.64 bps/Hz to 2.03 bps/Hz at a diminishing speed, while that of the FPA approach is 1.17 bps/Hz, which indicates the position optimization achieves $73\%$ performance improvement. For $K = 5$, the minimum user rate of the fluid antenna-enhanced system rise from 2.49 bps/Hz to 2.73 bps/Hz, while the FPA approach maintains a rate of 1.40 bps/Hz, resulting in a $95\%$ performance gain. Notably, for $K =3$, the minimum user rate remains approximately 3.46 bps/Hz within the Rician factor range of 2.5 to 6, which indicates it has converged. Compared to the FPA systems, FAS achieves a $54\%$ improvement. 
These results indicate that expanding the feasible region allows the system to better exploit spatial DoFs, thereby enhancing spatial diversity and mitigating multiuser interference. However, as the movable region becomes sufficiently large, the FAS has fully utilized the spatial variations in the channel, leading to diminishing returns in performance improvement. Furthermore, for a lower number of users, the movable region required for convergence of the minimum user rate is smaller, suggesting that a more compact moving region is sufficient to achieve substantial performance gains with fewer users.

\section{Conclusion} \label{sec7}
This paper investigated the two-timescale uplink transmission design for multiuser MIMO-FAS with imperfect CSI. Specifically, the BS performed beamforming to adapt to fast-changing instantaneous CSI in the small timescale, while the fluid antenna positions were optimized based on statistical CSI in the large timescale. First, we applied the Rician channel model and estimated the channel at the BS utilizing the LMMSE estimator. The economical MRC detector was then used, and a closed-form expression of the achievable user rate was derived. The problem was formulated as the maximization of the minimum achievable rate for all users, subject to the minimum spacing constraint and the feasible moving region constraint. Accordingly, a simple but computational complex GA-based method was proposed to tackle this non-convex problem. In order to reduce the complexity of numerical computation, an accelerated gradient ascent algorithm was introduced. Numerical results validated the potential of two algorithms for multiuser MIMO-FAS compared with traditional FPA schemes. Additionally, we examined the impact of key parameters on system performance.

\appendices
\section{} \label{appendix_A}
Using the observation signal $\mathbf{y}_p^k$, the LMMSE estimate of the channel vector $\mathbf{h}_k$ is given by \cite{ref38}
\begin{equation}
	\label{eq:41}
	\hat{\mathbf{h}}_k = \mathbb{E}\left\{\mathbf{h}_k\right\} + \text{Cov}\left\{\mathbf{h}_k,\mathbf{y}_p^k\right\}\text{Cov}^{-1}\left\{\mathbf{y}_p^k,\mathbf{y}_p^k\right\}\left(\mathbf{y}_p^k-\mathbb{E}\left\{\mathbf{y}_p^k\right\}\right).
\end{equation}

The mean vectors and covariance matrices required in (\ref{eq:41}) for the channel vector $\mathbf{h}_k$ and the observation signal $\mathbf{y}_p^k$ are calculated step by step.

First, recalling the definition of $\mathbf{y}_p^k$, where $\tilde{\mathbf{h}}_k$ and $\mathbf{N}$ are independent with zero-mean vector, the expectation term in (\ref{eq:41}) is given by
\begin{align}
	\mathbb{E}\left\{\mathbf{y}_p^k\right\} &= \mathbb{E}\left\{\mathbf{h}_k\right\} + \frac{1}{\sqrt{\tau p}}\mathbb{E}\left\{\mathbf{N}\right\}\mathbf{s}_k = \mathbb{E}\left\{\mathbf{h}_k\right\}, 
	\label{eq:42}\\
	\mathbb{E}\left\{\mathbf{h}_k\right\} &= \sqrt{c_k \varepsilon_k}\bar{\mathbf{h}}_k.
	\label{eq:43}
\end{align}

Then, the covariance matrices are given by
\begin{align}
	&\text{Cov} \left\{ \mathbf{h}_k,\mathbf{h}_k\right\}\nonumber \\
	&= \mathbb{E} \left\{ \left(\mathbf{h}_k - \mathbb{E} \left\{ \mathbf{h}_k\right\}\right) \left(\mathbf{h}_k - \mathbb{E} \left\{ \mathbf{h}_k\right\}\right)^H \right\} \nonumber \\			
	&= c_k \mathbb{E} \left\{ \tilde{\mathbf{h}}_k \tilde{\mathbf{h}}_k^H \right\}\nonumber\\
	&= c_k\mathbf{I}_M, 
	\label{eq:44}\\
	&\text{Cov} \left\{ \mathbf{h}_k, \mathbf{y}_p^k\right\}\nonumber \\
	&= \mathbb{E} \left\{ \left(\mathbf{h}_k - \mathbb{E} \left\{ \mathbf{h}_k\right\}\right)
	\left(\mathbf{y}_p^k - \mathbb{E}\left\{ \mathbf{y}_p^k\right\}\right)^H\right\}\nonumber \\
	&= \mathbb{E} \left\{ \left(\mathbf{h}_k - \mathbb{E} \left\{ \mathbf{h}_k\right\}\right) \left(\mathbf{h}_k + \frac{\mathbf{N} \mathbf{s}_k}{\sqrt{\tau p}} - \mathbb{E}\left\{ \mathbf{h}_k\right\}\right)^H\right\}\nonumber\\
	&= \mathbb{E} \bigg\{ \left(\mathbf{h}_k - \mathbb{E} \left\{ \mathbf{h}_k\right\}\right) \left(\mathbf{h}_k - \mathbb{E} \left\{ \mathbf{h}_k\right\}\right)^H\nonumber \\
	&\quad- \left(\mathbf{h}_k - \mathbb{E} \left\{ \mathbf{h}_k\right\}\right)\frac{\mathbf{s}_k^H\mathbf{N}^H}{\sqrt{\tau p}}   \bigg\}\nonumber \\
	&= \mathbb{E} \left\{ \left(\mathbf{h}_k - \mathbb{E} \left\{ \mathbf{h}_k\right\}\right) \left(\mathbf{h}_k - \mathbb{E} \left\{ \mathbf{h}_k\right\}\right)^H\right\}\nonumber \\
	&= \text{Cov} \left\{ \mathbf{h}_k, \mathbf{h}_k\right\},
	\label{eq:45}\\
	&\text{Cov} \left\{ \mathbf{y}_p^k,\mathbf{h}_k\right\}\nonumber\\ 	
	&= \left(\text{Cov} \left\{ \mathbf{h}_k, \mathbf{y}_p^k\right\} \right)^H \nonumber\\
	&= \text{Cov} \left\{ \mathbf{h}_k, \mathbf{h}_k\right\},
	\label{eq:46}\\
	&\text{Cov} \left\{ \mathbf{y}_p^k,\mathbf{y}_p^k\right\}\nonumber\\ 
	&= \mathbb{E} \left\{ \left(\mathbf{y}_p^k - \mathbb{E} \left\{ \mathbf{y}_p^k\right\}\right) \left(\mathbf{y}_p^k - \mathbb{E} \left\{ \mathbf{y}_p^k\right\}\right)^H \right\}\nonumber \\			
	&= \text{Cov} \left\{ \mathbf{h}_k,\mathbf{h}_k\right\} + \frac{1}{\tau p}\mathbb{E}\left\{\mathbf{N}\mathbf{s}_k\mathbf{s}_k^H \mathbf{N}^H\right\}\nonumber\\
	&= \text{Cov} \left\{ \mathbf{h}_k,\mathbf{h}_k\right\} + \frac{\sigma^2}{\tau p}\mathbf{I}_M\nonumber\\
	&= \left(c_k + \frac{\sigma^2}{\tau p}\right)\mathbf{I}_M.
	\label{eq:47}
\end{align}

Hence, the LMMSE channel estimate is calculated as
\begin{align}
	\label{eq:48}
	&\hat{\mathbf{h}}_k  \nonumber\\
	&= \sqrt{c_k\varepsilon_k}\bar{\mathbf{h}}_k + c_k\left(c_k+\frac{\sigma^2}{\tau p}\right)^{-1}\mathbf{I}_M\left(\mathbf{y}_p^k-\sqrt{c_k\varepsilon_k}\bar{\mathbf{h}}_k\right) \nonumber\\
	&= \frac{c_k}{c_k + \frac{\sigma^2}{\tau p}}\mathbf{y}_p^k + \sqrt{c_k\varepsilon_k}\bar{\mathbf{h}}_k -  \frac{c_k}{c_k + \frac{\sigma^2}{\tau p}}\sqrt{c_k\varepsilon_k}\bar{\mathbf{h}}_k \nonumber\\
	&= \frac{c_k}{c_k + \frac{\sigma^2}{\tau p}}\mathbf{y}_p^k + \frac{\frac{\sigma^2}{\tau p}}{c_k + \frac{\sigma^2}{\tau p}}\sqrt{c_k\varepsilon_k}\bar{\mathbf{h}}_k \nonumber\\
	&\triangleq a_k\mathbf{y}_k + \mathbf{b}_k \nonumber\\
	&= \frac{c_k}{c_k + \frac{\sigma^2}{\tau p}}\left(\mathbf{h}_k + \frac{1}{\tau p}\mathbf{N}\mathbf{s}_k\right) + \frac{\frac{\sigma^2}{\tau p}}{c_k + \frac{\sigma^2}{\tau p}}\sqrt{c_k\varepsilon_k}\bar{\mathbf{h}}_k \nonumber\\
	&= \frac{c_k}{c_k + \frac{\sigma^2}{\tau p}}\left(\sqrt{c_k\varepsilon_k}\bar{\mathbf{h}}_k + \sqrt{c_k}\tilde{\mathbf{h}}_k + \frac{1}{\tau p}\mathbf{N}\mathbf{s}_k\right) \nonumber\\
	&\quad + \frac{\frac{\sigma^2}{\tau p}}{c_k + \frac{\sigma^2}{\tau p}}\sqrt{c_k\varepsilon_k}\bar{\mathbf{h}}_k \nonumber\\
	&= \sqrt{c_k\varepsilon_k}\bar{\mathbf{h}}_k + \frac{c_k}{c_k+\frac{\sigma^2}{\tau p}}\sqrt{c_k}\tilde{\mathbf{h}}_k + \frac{c_k}{c_k+\frac{\sigma^2}{\tau p}}\frac{1}{\sqrt{\tau p}}\mathbf{N}\mathbf{s}_k \nonumber\\
	&= \sqrt{c_k\varepsilon_k}\bar{\mathbf{h}}_k + a_k\sqrt{c_k}\tilde{\mathbf{h}}_k + \frac{a_k}{\sqrt{\tau p}}\mathbf{N}\mathbf{s}_k. 
\end{align}

\section{} \label{appendix_B}
To begin with, we present some properties which will be useful in the following derivations.
\begin{lemma} 
	\label{lem-2}
	Let $\mathbf{X}\in \mathbb{C}^{m \times n}$, $m, n \ge 1$, be a matrix whose elements are i.i.d. random variables with zero mean and $v_x$ variance. Consider a deterministic matrix $A \in \mathbb{C}^{n \times n}$. Then, the following holds
	\begin{equation}
		\label{eq:49}
		\mathbb{E}\left\{\mathbf{X}\mathbf{A}\mathbf{X}^H\right\}=v_x\mathrm{Tr}\left\{\mathbf{A}\right\}\mathbf{I}_m.
	\end{equation}
\end{lemma}
\itshape {Proof:}  \upshape Consider the matrix $\mathbf{X}\mathbf{W}\mathbf{A}^H$. Due to the entries of $\mathbf{X}$ are i.i.d with zero mean, the expectation of the $(i,j)$-th entry, where $i\neq j$, is given by
\begin{align}
	\label{eq:50}
	\left[\mathbb{E}\left\{\mathbf{X}\mathbf{A}\mathbf{X}^H\right\}\right]_{i, j} 
	&=\mathbb{E}\left\{\sum_{l=1}^{n}\sum_{k=1}^{n}[\mathbf{X}]_{i, k}[\mathbf{A}]_{k, l}\left[\mathbf{X}^H\right]_{l, j}\right\} \nonumber\\
	&=\sum_{l=1}^{n}\sum_{k=1}^{n}\mathbb{E}\left\{[\mathbf{X}]_{i, k}[\mathbf{X}^*]_{j, l}\right\}[\mathbf{A}]_{k, l} = 0.
\end{align}

Similarly, owing to the variance of $\mathbf{X}$ elements, the expectation of the diagonal entry is 
\begin{align}
	\label{eq:51}
	\left[\mathbb{E}\left\{\mathbf{X}\mathbf{A}\mathbf{X}^H\right\}\right]_{i, i}
	&=\sum_{l=1}^{n}\sum_{k=1}^{n}\mathbb{E}\left\{[\mathbf{X}]_{i, k}[\mathbf{X}]^*_{i, l}\right\}[\mathbf{A}]_{k, l} \nonumber\\
	&=\sum_{k=1}^{n}\mathbb{E}\left\{\left|[\mathbf{X}]_{i, k}\right|^2\right\}[\mathbf{A}]_{k, k}=v_x \mathrm{Tr}\left\{\mathbf{A}\right\}.
\end{align}

Thus, the expected value of $\mathbf{X}\mathbf{A}\mathbf{X}^H$ is a diagonal matrix, with each diagonal entry equal to  $v_x \mathrm{Tr}\left\{\mathbf{A}\right\}$. This completes the proof. 

\begin{lemma} 
	\label{lem-3}
	Consider vectors $\mathbf{u}_1, \mathbf{u}_2 \in \mathbb{C}^{N\times 1}$, Then we have
	\begin{align}
		&\mathbb{E}\left\{\mathbf{u}_1^H\tilde{\mathbf{h}}_k\mathbf{u}_2^H\tilde{\mathbf{h}}_k\right\}
		=\mathbb{E}\left\{\mathrm{Re}\left\{\mathbf{u}_1^H\tilde{\mathbf{h}}_k\mathbf{u}_2^H\tilde{\mathbf{h}}_k\right\}\right\}=0, \label{eq:52}\\
		&\mathbb{E}\left\{\left\|\tilde{\mathbf{h}}_k\right\|^4\right\} = \mathrm{Tr}\left\{\mathbb{E}\left\{\tilde{\mathbf{h}}_k\tilde{\mathbf{h}}_k^H\tilde{\mathbf{h}}_k\tilde{\mathbf{h}}_k^H\right\}\right\} = M^2+M. \label{eq:53}
	\end{align}
\end{lemma}

\itshape {Proof:}  \upshape Let $v=v_r+jv_i$ represents a complex random variable, where $v_r$ and $v_i$ are independent and normally distributed with zero mean and variance $\frac{1}{2}$, i.e., $v_r, v_i ~ \mathcal{N}(0,\frac{1}{2})$. From this, we derive
\begin{align}
	&\mathbb{E}\left\{v^2\right\}=\mathbb{E}\left\{v_r^2-v_i^2+2jv_rv_i\right\}=0, \label{eq:54}\\
	&\mathbb{E}\left\{\mathrm{Re}\left\{v^2\right\}\right\} = \mathbb{E}\left\{v_r^2\right\} - \mathbb{E}\left\{v_i^2\right\} =0. \label{eq:55}
\end{align}

The entries of $\tilde{\mathbf{h}}_k$ are i.i.d, each following the same distribution as $v$. Then, we can express the expectation of (\ref{eq:52}) as
\begin{align}
	\label{eq:56}
	&\mathbb{E}\left\{\mathbf{u}_1^H\tilde{\mathbf{h}}_k\mathbf{u}_2^H\tilde{\mathbf{h}}_k\right\} \nonumber\\
	&=\mathbb{E}\left\{\sum_{n_1=1}^{M}\sum_{n_2=1}^{M}[\mathbf{u}_1^H]_{n_1}[\mathbf{u}_2^H]_{n2}[\tilde{\mathbf{h}}_k]_{n1}[\tilde{\mathbf{h}}_k]_{n2}\right\}.
\end{align}

When $n1 \neq n2$, the expectation $\mathbb{E}\left\{[\mathbf{u}_1^H]_{n_1}[\mathbf{u}_2^H]_{n2}[\tilde{\mathbf{h}}_k]_{n1}[\tilde{\mathbf{h}}_k]_{n2}\right\}$ vanishes due to the independence and zero mean of the elements of $\tilde{\mathbf{h}}_k$.  
When $n1 = n2$, the expectation becomes zero according to (\ref{eq:54}). Hence, (\ref{eq:52}) is verified. 

Next, consider the expectation of $\left\|\tilde{\mathbf{h}}_k\right\|^4$, which can be expressed as 
\begin{align}
	\label{eq:57}
	\mathbb{E}\left\{\left\|\tilde{\mathbf{h}}_k\right\|^4\right\} 
	&=\mathbb{E}\left\{\left(\sum_{i=1}^{M}\left|[\tilde{\mathbf{h}}_k]_i\right|^2\right)^2\right\} \nonumber\\
	&=\mathbb{E}\left\{\sum_{i=1}^{M}\left|[\tilde{\mathbf{h}}_k]_i\right|^4\right\} + \mathbb{E}\left\{\sum_{i \neq j}\left|[\tilde{\mathbf{h}}_k]_i\right|^2\left|[\tilde{\mathbf{h}}_k]_j\right|^2\right\} \nonumber\\
	&=2M+M(M-1) =M^2+M.
\end{align}

Therefore, (\ref{eq:53}) is established, completing the proof.

\subsection {Signal Term and Noise Term}
Based on the estimated channel $\hat{\mathbf{h}}_k$ in (\ref{eq:12}), the estimation error $\mathbf{e}_k = \mathbf{h}_k-\hat{\mathbf{h}}_k$ has zero mean. Then, according to the orthogonality of the LMMSE estimator, we can obtain $\mathbb{E}\left\{\mathbf{e}_k\left(\mathbf{y}_p^k\right)^H\right\}=\mathbf{0}$, and it follows that $\mathbb{E}\left\{\hat{\mathbf{h}}_k^H \mathbf{e}_k\right\}=\mathbf{0}$. Therefore, We have
\begin{equation}
	\label{eq:58}
	\mathbb{E}\left\{\hat{\mathbf{h}}_k^H\mathbf{h}_k\right\}=\mathbb{E}\left\{\hat{\mathbf{h}}_k^H\hat{\mathbf{h}}_k\right\}+\mathbb{E}\left\{\hat{\mathbf{h}}_k^H \mathbf{e}_k\right\}=\mathbb{E}\left\{\|\hat{\mathbf{h}}_k\|^2\right\}.
\end{equation}

Hence, the signal power can be expressed as
\begin{equation}
	\label{eq:59}
	E_k^{signal}= \left|\mathbb{E}\left\{\hat{\mathbf{h}}_k^H\mathbf{h}_k\right\}\right|^2 = \left(\mathbb{E}\left\{\|\hat{\mathbf{h}}_k\|^2\right\}\right)^2 = \left(E_k^{noise}\right)^2.
\end{equation}

Then, the noise term can be expanded as
\begin{equation}
	\begin{aligned}
		\label{eq:60}
		E_k^{noise}&=\mathbb{E}\left\{\|\hat{\mathbf{h}}_k\|^2\right\} = \mathbb{E}\left\{\hat{\mathbf{h}}_k^H\mathbf{h}_k\right\} \\
		&=\mathbb{E}\big\{(\sqrt{c_k\varepsilon_k}\bar{\mathbf{h}}_k+a_k\sqrt{c_k}\tilde{\mathbf{h}}_k+\frac{a_k}{\sqrt{\tau p}}\mathbf{N}\mathbf{s}_k)^H\\
		&\quad (\sqrt{c_k\varepsilon_k}\bar{\mathbf{h}}_k+\sqrt{c_k}\tilde{\mathbf{h}}_k)\big\} \\
		&=c_k\varepsilon_k\bar{\mathbf{h}}_k^H\bar{\mathbf{h}} +a_kc_k\mathbb{E}\left\{\tilde{\mathbf{h}}_k^H\tilde{\mathbf{h}}_k\right\} \\
		&=Mc_k(\varepsilon_k+a_k).
	\end{aligned}
\end{equation}
\subsection {Signal Leakage}
The signal leakage can be expressed as $E_k^{leak} = \mathbb{E}\left\{\left|\hat{\mathbf{h}}_k^H\mathbf{h}_k\right|^2\right\} - \left|\mathbb{E}\left\{\hat{\mathbf{h}}_k^H \mathbf{h}_k \right\}\right|^2$ in (\ref{eq:17}), where $\mathbb{E}\left\{\hat{\mathbf{h}}_k^H \mathbf{h}_k \right\}$ is given in (\ref{eq:60}), and $\mathbb{E}\left\{\left|\hat{\mathbf{h}}_k^H\mathbf{h}_k\right|^2\right\}$ will be derived by exploiting the independence between the channel and the noise, as follows
\begin{equation}
	\begin{aligned}
		\label{eq:61}
		&\mathbb{E}\bigg\{\left|\hat{\mathbf{h}}_k^H\mathbf{h}_k\right|^2\bigg\}\\
		&= \mathbb{E}\bigg\{\bigg|(\sqrt{c_k\varepsilon_k}\bar{\mathbf{h}}_k+a_k\sqrt{c_k}\tilde{\mathbf{h}}_k+\frac{a_k}{\sqrt{\tau p}}\mathbf{N}\mathbf{s}_k)^H \\ 
		&\quad\quad\quad(\sqrt{c_k\varepsilon_k}\bar{\mathbf{h}}_k+\sqrt{c_k}\tilde{\mathbf{h}}_k)\bigg|^2\bigg\} \\
		&=\mathbb{E}\bigg\{\bigg|c_k\varepsilon_k\bar{\mathbf{h}}_k^H\bar{\mathbf{h}}_k
		+c_k\sqrt{\varepsilon_k}\bar{\mathbf{h}}_k^H\tilde{\mathbf{h}}_k \\
		&\quad +a_kc_k\sqrt{\varepsilon_k}\tilde{\mathbf{h}}_k^H\bar{\mathbf{h}}_k
		+a_kc_k\tilde{\mathbf{h}}_k^H\tilde{\mathbf{h}}_k \\
		&\quad +\frac{a_k}{\sqrt{\tau p}}\sqrt{c_k\varepsilon_k}\mathbf{s}_k^H\mathbf{N}^H\bar{\mathbf{h}}_k
		+\frac{a_k}{\sqrt{\tau p}}\sqrt{c_k}\mathbf{s}_k^H\mathbf{N}^H\tilde{\mathbf{h}}_k\bigg|^2\bigg\}\\
		&=\left|c_k\varepsilon_k\bar{\mathbf{h}}_k^H\bar{\mathbf{h}}_k\right|^2 +\mathbb{E}\left\{\left|c_k\sqrt{\varepsilon_k}\bar{\mathbf{h}}_k^H\tilde{\mathbf{h}}_k\right|^2\right\} \\
		&\quad +\mathbb{E}\left\{\left|a_kc_k\sqrt{\varepsilon_k}\tilde{\mathbf{h}}_k^H\bar{\mathbf{h}}_k\right|^2\right\} 
		+\mathbb{E}\left\{\left|a_kc_k\tilde{\mathbf{h}}_k^H\tilde{\mathbf{h}}_k\right|^2\right\} \\
		&\quad +\mathbb{E}\left\{\left|\frac{a_k}{\sqrt{\tau p}}\sqrt{c_k\varepsilon_k}\mathbf{s}_k^H\mathbf{N}^H\bar{\mathbf{h}}_k\right|^2\right\}\\
		&\quad+\mathbb{E}\left\{\left|\frac{a_k}{\sqrt{\tau p}}\sqrt{c_k}\mathbf{s}_k^H\mathbf{N}^H\tilde{\mathbf{h}}_k\right|^2\right\}\\
		&\quad+2\mathrm{Re}
		\left\{\mathbb{E}\left\{c_k^2a_k\varepsilon_k\bar{\mathbf{h}}_k^H\bar{\mathbf{h}}_k\tilde{\mathbf{h}}_k^H\tilde{\mathbf{h}}_k\right\}\right\}.
	\end{aligned}
\end{equation}

Then, the seven terms in (\ref{eq:61}) will be calculated as follows
\begin{align}
	&\left|c_k\varepsilon_k\bar{\mathbf{h}}_k^H\bar{\mathbf{h}}_k\right|^2=M^2c_k^2\varepsilon_k^2, 
	\label{eq:62}\\
	&\mathbb{E}\left\{\left|c_k\sqrt{\varepsilon_k}\bar{\mathbf{h}}_k^H\tilde{\mathbf{h}}_k\right|^2\right\}=c_k^2\varepsilon_k\bar{\mathbf{h}}_k^H\mathbb{E}\left\{\tilde{\mathbf{h}}_k\tilde{\mathbf{h}}_k^H\right\}\bar{\mathbf{h}}_k=Mc_k^2\varepsilon_k, 
	\label{eq:63}\\
	&\mathbb{E}\left\{\left|a_kc_k\sqrt{\varepsilon_k}\tilde{\mathbf{h}}_k^H\bar{\mathbf{h}}_k\right|^2\right\}=a_k^2c_k^2\varepsilon_k\mathbb{E}\left\{\tilde{\mathbf{h}}_k^H\bar{\mathbf{h}}_k\bar{\mathbf{h}}_k^H\tilde{\mathbf{h}}_k\right\} \nonumber\\
	&\overset{(a)}{=} a_k^2c_k^2\varepsilon_k\mathrm{Tr}\left\{\bar{\mathbf{h}}_k\bar{\mathbf{h}}_k^H\right\}
	=Ma_k^2c_k^2\varepsilon_k, 
	\label{eq:64}\\
	&\mathbb{E}\left\{\left|a_kc_k\tilde{\mathbf{h}}_k^H\tilde{\mathbf{h}}_k\right|^2\right\} \nonumber\\
	&=a_k^2c_k^2\mathbb{E}\left\{\tilde{\mathbf{h}}_k^H\tilde{\mathbf{h}}_k\tilde{\mathbf{h}}_k^H\tilde{\mathbf{h}}_k\right\}
	\overset{(b)}{=}M(M+1)a_k^2c_k^2,
	\label{eq:65}\\
	&\mathbb{E}\left\{\left|\frac{a_k}{\sqrt{\tau p}}\sqrt{c_k\varepsilon_k}\mathbf{s}_k^H\mathbf{N}^H\bar{\mathbf{h}}_k\right|^2\right\} \nonumber\\
	&=\frac{1}{\tau p}a_k^2c_k\varepsilon_k\mathbb{E}\left\{\mathbf{s}_k^H\mathbf{N}^H\bar{\mathbf{h}}_k\bar{\mathbf{h}}_k^H\mathbf{N}\mathbf{s}_k\right\} \nonumber\\
	&=\frac{1}{\tau p}a_k^2c_k\varepsilon_k\mathbf{s}_k^H\mathbb{E}\left\{\mathbf{N}^H\bar{\mathbf{h}}_k\bar{\mathbf{h}}_k^H\mathbf{N}\right\}\mathbf{s}_k \nonumber\\
	&\overset{(c)}{=}\frac{\sigma^2}{\tau p}a_k^2c_k\varepsilon_k\mathbf{s}_k^H\mathrm{Tr}\left\{\bar{\mathbf{h}}_k\bar{\mathbf{h}}_k^H\right\}\mathbf{s}_k \nonumber\\
	&=\frac{\sigma^2}{\tau p}Ma_k^2c_k\varepsilon_k,
	\label{eq:66} \\
	&\mathbb{E}\left\{\left|\frac{a_k}{\sqrt{\tau p}}\sqrt{c_k}\mathbf{s}_k^H\mathbf{N}^H\tilde{\mathbf{h}}_k\right|^2\right\} \nonumber\\
	&=\frac{1}{\tau p}a_k^2c_k\mathbb{E}\left\{\mathbf{s}_k^H\mathbf{N}^H\tilde{\mathbf{h}}_k\tilde{\mathbf{h}}_k^H\mathbf{N}\mathbf{s}_k\right\} \nonumber\\
	&=\frac{1}{\tau p}a_k^2c_k\mathbf{s}_k^H\mathbb{E}\left\{\mathbf{N}^H\mathbb{E}\left\{\tilde{\mathbf{h}}_k\tilde{\mathbf{h}}_k^H\right\}\mathbf{N}\right\}\mathbf{s}_k \nonumber\\
	&=\frac{\sigma^2}{\tau p}a_k^2c_k,
	\label{eq:67} \\
	&2\mathrm{Re}
	\left\{\mathbb{E}\left\{c_k^2a_k\varepsilon_k\bar{\mathbf{h}}_k^H\bar{\mathbf{h}}_k\tilde{\mathbf{h}}_k^H\tilde{\mathbf{h}}_k\right\}\right\} \nonumber\\
	&=2c_k^2a_k\varepsilon_k\mathrm{Re}\left\{\bar{\mathbf{h}}_k^H\bar{\mathbf{h}}_k\mathbb{E}\left\{\tilde{\mathbf{h}}_k^H\tilde{\mathbf{h}}_k\right\}\right\} \nonumber\\
	&=2M^2c_k^2a_k\varepsilon_k,
	\label{eq:68}
\end{align}
where (a) and (c) apply Lemma \ref{lem-2}, and (b) applies Lemma \ref{lem-3}. Thus, we can obtain the $\mathbb{E}\bigg\{\left|\hat{\mathbf{h}}_k^H\mathbf{h}_k\right|^2\bigg\}$ by combining (\ref{eq:62})-(\ref{eq:68}) with (\ref{eq:61}). With the aid of (\ref{eq:60}), we can complete the calculation of the signal leakage term $E_k^{leak}$ using (\ref{eq:17}).
\subsection {Interference Term}
To begin, we obtain the interference term by breaking it down as follows
\begin{equation}
	\begin{aligned}
		\label{eq:69}
		&I_{ki} \triangleq \mathbb{E}\left\{\left|\hat{\mathbf{h}}_k^H \mathbf{h}_i\right|^2\right\} \\
		&=\mathbb{E}\bigg\{\bigg|(\sqrt{c_k\varepsilon_k}\bar{\mathbf{h}}_k
		+a_k\sqrt{c_k}\tilde{\mathbf{h}}_k+\frac{a_k}{\sqrt{\tau p}}\mathbf{N}\mathbf{s}_k)^H \\ &\qquad\qquad\qquad\qquad\qquad(\sqrt{c_i\varepsilon_i}\bar{\mathbf{h}}_i
		+\sqrt{c_i}\tilde{\mathbf{h}}_i)\bigg|^2\bigg\} \\
		&=\mathbb{E}\bigg\{\bigg|\sqrt{c_kc_i\varepsilon_k\varepsilon_i}\bar{\mathbf{h}}_k^H\bar{\mathbf{h}}_i
		+\sqrt{c_kc_i\varepsilon_k}\bar{\mathbf{h}}_k^H\tilde{\mathbf{h}}_i \\
		&\quad+a_k\sqrt{c_kc_i\varepsilon_i}\tilde{\mathbf{h}}_k^H\bar{\mathbf{h}}_i
		+a_k\sqrt{c_kc_i}\tilde{\mathbf{h}}_k^H\tilde{\mathbf{h}}_i \\
		&\quad+\frac{a_k}{\sqrt{\tau p}}\sqrt{c_i\varepsilon_i}\mathbf{s}_k^H\mathbf{N}^H\bar{\mathbf{h}}_i
		+\frac{a_k}{\sqrt{\tau p}}\sqrt{c_i}\mathbf{s}_k^H\mathbf{N}^H\tilde{\mathbf{h}}_i\bigg|^2\bigg\} \\
		&=\left|\sqrt{c_kc_i\varepsilon_k\varepsilon_i}\bar{\mathbf{h}}_k^H\bar{\mathbf{h}}_i\right|^2
		+\mathbb{E}\left\{\left|\sqrt{c_kc_i\varepsilon_k}\bar{\mathbf{h}}_k^H\tilde{\mathbf{h}}_i\right|^2\right\} \\
		&\quad+\mathbb{E}\left\{\left|a_k\sqrt{c_kc_i\varepsilon_i}\tilde{\mathbf{h}}_k^H\bar{\mathbf{h}}_i\right|^2\right\}
		+\mathbb{E}\left\{\left|a_k\sqrt{c_kc_i}\tilde{\mathbf{h}}_k^H\tilde{\mathbf{h}}_i\right|^2\right\} \\
		&\quad+\mathbb{E}\left\{\left|\frac{a_k}{\sqrt{\tau p}}\sqrt{c_i\varepsilon_i}\mathbf{s}_k^H\mathbf{N}^H\bar{\mathbf{h}}_i\right|^2\right\} \\
		&\quad+\mathbb{E}\left\{\left|\frac{a_k}{\sqrt{\tau p}}\sqrt{c_i}\mathbf{s}_k^H\mathbf{N}^H\tilde{\mathbf{h}}_i\right|^2\right\}.
	\end{aligned}
\end{equation}

Then, the six term in (\ref{eq:69}) will be given by
\begin{align}
	&\left|\sqrt{c_kc_i\varepsilon_k\varepsilon_i}\bar{\mathbf{h}}_k^H\bar{\mathbf{h}}_i\right|^2
	=c_kc_i\varepsilon_k\varepsilon_i\left|\bar{\mathbf{h}}_k^H\bar{\mathbf{h}}_i\right|^2
	=c_kc_i\varepsilon_k\varepsilon_i\left|f_k(\tilde{\mathbf{t}})\right|^2,
	\label{eq:70}\\
	&\mathbb{E}\left\{\left|\sqrt{c_kc_i\varepsilon_k}\bar{\mathbf{h}}_k^H\tilde{\mathbf{h}}_i\right|^2\right\}
	=c_kc_i\varepsilon_k\bar{\mathbf{h}}_k^H\mathbb{E}\left\{\tilde{\mathbf{h}}_i\tilde{\mathbf{h}}_i^H\right\}\bar{\mathbf{h}}_k \nonumber\\
	&=Mc_kc_i\varepsilon_k,
	\label{eq:71}\\
	&\mathbb{E}\left\{\left|a_k\sqrt{c_kc_i\varepsilon_i}\tilde{\mathbf{h}}_k^H\bar{\mathbf{h}}_i\right|^2\right\}
	=a_k^2c_kc_i\varepsilon_i\mathbb{E}\left\{\tilde{\mathbf{h}}_k^H\bar{\mathbf{h}}_i\bar{\mathbf{h}}_i^H\tilde{\mathbf{h}}_k\right\} \nonumber\\
	&\overset{(d)}{=}a_k^2c_kc_i\varepsilon_i\mathrm{Tr}\left\{\bar{\mathbf{h}}_i\bar{\mathbf{h}}_i^H\right\}
	=Ma_k^2c_kc_i\varepsilon_i,
	\label{eq:72}\\
	&\mathbb{E}\left\{\left|a_k\sqrt{c_kc_i}\tilde{\mathbf{h}}_k^H\tilde{\mathbf{h}}_i\right|^2\right\} \nonumber\\
	&=a_k^2c_kc_i\mathbb{E}\left\{\tilde{\mathbf{h}}_k^H\mathbb{E}\left\{\tilde{\mathbf{h}}_i\tilde{\mathbf{h}}_i^H\right\}\tilde{\mathbf{h}}_k\right\}
	=Ma_k^2c_kc_i,
	\label{eq:73}\\
	&\mathbb{E}\left\{\left|\frac{a_k}{\tau p}\sqrt{c_i\varepsilon_i}\mathbf{s}_k^H\mathbf{N}^H\bar{\mathbf{h}}_i\right|^2\right\} \nonumber\\
	&=\frac{1}{\tau p}a_k^2c_i\varepsilon_i\mathbf{s}_k^H\mathbb{E}\left\{\mathbf{N}^H\bar{\mathbf{h}}_i\bar{\mathbf{h}}_i^H\mathbf{N}\right\}\mathbf{s}_k \nonumber\\
	&\overset{(e)}{=}\frac{\sigma^2}{\tau p}a_k^2c_i\varepsilon_i\mathbf{s}_k^H\mathrm{Tr}\left\{\bar{\mathbf{h}}_i\bar{\mathbf{h}}_i^H\right\}\mathbf{s}_k \nonumber\\
	&=\frac{\sigma^2}{\tau p}Ma_k^2c_i\varepsilon_i,
	\label{eq:74}\\
	&\mathbb{E}\left\{\left|\frac{a_k}{\tau p}\sqrt{c_i}\mathbf{s}_k^H\mathbf{N}^H\tilde{\mathbf{h}}_i\right|^2\right\} \nonumber\\
	&=\frac{1}{\tau p}a_k^2c_i\mathbf{s}_k^H\mathbb{E}\left\{\mathbf{N}^H\mathbb{E}\left\{\tilde{\mathbf{h}}_i\tilde{\mathbf{h}}_i^H\right\}\mathbf{N}\right\}\mathbf{s}_k = \frac{\sigma^2}{\tau p}a_k^2c_i,
	\label{eq:75}
\end{align}
where (d) and (e) apply Lemma \ref{lem-2}. Consequently, the calculation is finalized, and the interference term is obtained by integrating (\ref{eq:70}) - (\ref{eq:75}) with (\ref{eq:69}).

\bibliographystyle{IEEEtran}
\bibliography{IEEEabrv,reference}

\begin{thebibliography}{10}
\providecommand{\url}[1]{#1}
\csname url@samestyle\endcsname
\providecommand{\newblock}{\relax}
\providecommand{\bibinfo}[2]{#2}
\providecommand{\BIBentrySTDinterwordspacing}{\spaceskip=0pt\relax}
\providecommand{\BIBentryALTinterwordstretchfactor}{4}
\providecommand{\BIBentryALTinterwordspacing}{\spaceskip=\fontdimen2\font plus
\BIBentryALTinterwordstretchfactor\fontdimen3\font minus
  \fontdimen4\font\relax}
\providecommand{\BIBforeignlanguage}[2]{{%
\expandafter\ifx\csname l@#1\endcsname\relax
\typeout{** WARNING: IEEEtran.bst: No hyphenation pattern has been}%
\typeout{** loaded for the language `#1'. Using the pattern for}%
\typeout{** the default language instead.}%
\else
\language=\csname l@#1\endcsname
\fi
#2}}
\providecommand{\BIBdecl}{\relax}
\BIBdecl

\bibitem{ref1}
A.~Fayad, T.~Cinkler, and J.~Rak, ``Toward 6{G} optical fronthaul: A survey on
  enabling technologies and research perspectives,'' \emph{IEEE Commun. Surv.
  Tutorials}, vol.~27, no.~1, pp. 629--666, Feb. 2024.

\bibitem{ref2}
L.~Zheng and D.~N.~C. Tse, ``Diversity and multiplexing: A fundamental tradeoff
  in multiple-antenna channels,'' \emph{IEEE Trans. Inf. Theory}, vol.~49,
  no.~5, pp. 1073--1096, May 2003.

\bibitem{ref3}
X.~Shao, Q.~Jiang, and R.~Zhang, ``6{D} movable antenna based on user
  distribution: Modeling and optimization,'' \emph{IEEE Trans. Wireless
  Commun.}, vol.~24, no.~1, pp. 355--370, Jan. 2025.

\bibitem{ref4}
E.~G. Larsson, O.~Edfors, F.~Tufvesson, and T.~L. Marzetta, ``Massive {MIMO}
  for next generation wireless systems,'' \emph{IEEE Commun. Mag.}, vol.~52,
  no.~2, pp. 186--195, Feb. 2014.

\bibitem{ref5}
L.~Jing, M.~Li, and R.~Murch, ``Compact pattern reconfigurable pixel antenna
  with diagonal pixel connections,'' \emph{IEEE Trans. Antennas Propag.},
  vol.~70, no.~10, pp. 8951--8961, Oct. 2022.

\bibitem{ref6}
S.~Dash, C.~Psomas, and I.~Krikidis, ``Selection of metallic liquid in sub-6
  {GH}z antenna design for 6{G} networks,'' \emph{Sci. Rep.}, vol.~13, no.~1,
  p. 20551, Nov. 2023.

\bibitem{ref8}
S.~Basbug, ``Design and synthesis of antenna array with movable elements along
  semicircular paths,'' \emph{IEEE Antennas Wirel. Propag. Lett.}, vol.~16, pp.
  3059--3062, 2017.

\bibitem{ref9}
K.-K. Wong, K.-F. Tong, Y.~Shen, Y.~Chen, and Y.~Zhang, ``Bruce lee-inspired
  fluid antenna system: Six research topics and the potentials for 6{G},''
  \emph{Front. commun. netw}, vol.~3, p. 853416, Mar. 2022.

\bibitem{ref10}
K.-K. Wong and K.-F. Tong, ``Fluid antenna multiple access,'' \emph{IEEE Trans.
  Wireless Commun.}, vol.~21, no.~7, pp. 4801--4815, Jul. 2021.

\bibitem{ref11}
L.~Zhu, W.~Ma, and R.~Zhang, ``Movable-antenna array enhanced beamforming:
  Achieving full array gain with null steering,'' \emph{IEEE Commun. Lett.},
  vol.~27, no.~12, pp. 3340--3344, Dec. 2023.

\bibitem{ref12}
K.-K. Wong, A.~Shojaeifard, K.-F. Tong, and Y.~Zhang, ``Fluid antenna
  systems,'' \emph{IEEE Trans. Wireless Commun.}, vol.~20, no.~3, pp.
  1950--1962, Mar. 2020.

\bibitem{ref13}
K.~K. Wong, A.~Shojaeifard, K.-F. Tong, and Y.~Zhang, ``Performance limits of
  fluid antenna systems,'' \emph{IEEE Commun. Lett.}, vol.~24, no.~11, pp.
  2469--2472, Nov. 2020.

\bibitem{ref14}
P.~Mukherjee, C.~Psomas, and I.~Krikidis, ``On the level crossing rate of fluid
  antenna systems,'' in \emph{Proc. IEEE Int. Workshop Signal Process. Adv.
  Wireless Commun. (SPAWC)}, Jul. 2022, pp. 1--5.

\bibitem{ref15}
L.~Tlebaldiyeva, G.~Nauryzbayev, S.~Arzykulov, A.~Eltawil, and T.~Tsiftsis,
  ``Enhancing {Q}o{S} through fluid antenna systems over correlated nakagami-m
  fading channels,'' in \emph{Proc. IEEE Wireless Commun. Netw. Conf.
  (WCNC)}.\hskip 1em plus 0.5em minus 0.4em\relax IEEE, Apr. 2022, pp. 78--83.

\bibitem{ref16}
C.~Skouroumounis and I.~Krikidis, ``Fluid antenna with linear {MMSE} channel
  estimation for large-scale cellular networks,'' \emph{IEEE Trans. Commun.},
  vol.~71, no.~2, pp. 1112--1125, Feb. 2022.

\bibitem{ref17}
X.~Lai, T.~Wu, J.~Yao, C.~Pan, M.~Elkashlan, and K.-K. Wong, ``On performance
  of fluid antenna system using maximum ratio combining,'' \emph{IEEE Commun.
  Lett.}, vol.~28, no.~2, pp. 402--406, Feb. 2024.

\bibitem{ref18}
C.~Psomas and I.~Krikidis, ``Switched combining for reconfigurable fluid
  antenna systems,'' in \emph{Proc. IEEE Int. Workshop Signal Process. Adv.
  Wireless Commun. (SPAWC)}, Sep. 2024, pp. 281--285.

\bibitem{ref19}
L.~Zhu and K.-K. Wong, ``Historical review of fluid antenna and movable
  antenna,'' 2024, \textit{arXiv:2401.02362}.

\bibitem{ref20}
G.~Hu, Q.~Wu, D.~Xu, K.~Xu, J.~Si, Y.~Cai, and N.~Al-Dhahir, ``Movable
  antennas-assisted secure transmission without eavesdroppers' instantaneous
  {CSI},'' \emph{IEEE Trans. Mob. Comput.}, vol.~23, no.~12, pp.
  14\,263--14\,279, Dec. 2024.

\bibitem{ref21}
C.~Wang, G.~Li, H.~Zhang, K.-K. Wong, Z.~Li, D.~W.~K. Ng, and C.-B. Chae,
  ``Fluid antenna system liberating multiuser {MIMO} for {ISAC} via deep
  reinforcement learning,'' \emph{IEEE Trans. Wireless Commun.}, vol.~23,
  no.~9, pp. 10\,879--10\,894, Sep. 2024.

\bibitem{ref22}
J.~Zou, H.~Xu, C.~Wang, L.~Xu, S.~Sun, K.~Meng, C.~Masouros, and K.-K. Wong,
  ``Shifting the {ISAC} trade-off with fluid antenna systems,'' \emph{IEEE
  Wireless Commun. Lett.}, vol.~13, no.~12, pp. 3479--3483, Dec. 2024.

\bibitem{ref23}
F.~R. Ghadi, K.-K. Wong, W.~K. New, H.~Xu, R.~Murch, and Y.~Zhang, ``On
  performance of {RIS}-aided fluid antenna systems,'' \emph{IEEE Wireless
  Commun. Lett.}, vol.~13, no.~8, pp. 2175--2179, Aug. 2024.

\bibitem{ref24}
W.~K. New, K.-K. Wong, H.~Xu, K.-F. Tong, and C.-B. Chae, ``An
  information-theoretic characterization of {MIMO-FAS}: Optimization,
  diversity-multiplexing tradeoff and q-outage capacity,'' \emph{IEEE
  Transactions on Wireless Communications}, vol.~23, no.~6, pp. 5541--5556,
  Jun. 2023.

\bibitem{ref25}
W.~Ma, L.~Zhu, and R.~Zhang, ``{MIMO} capacity characterization for movable
  antenna systems,'' \emph{IEEE Trans. Wireless Commun.}, vol.~23, no.~4, pp.
  3392--3407, Apr. 2023.

\bibitem{ref26}
L.~Zhu, W.~Ma, B.~Ning, and R.~Zhang, ``Movable-antenna enhanced multiuser
  communication via antenna position optimization,'' \emph{IEEE Trans. Wireless
  Commun.}, vol.~23, no.~7, pp. 7214--7229, Jul. 2024.

\bibitem{ref27}
Z.~Xiao, X.~Pi, L.~Zhu, X.-G. Xia, and R.~Zhang, ``Multiuser communications
  with movable-antenna base station: Joint antenna positioning, receive
  combining, and power control,'' \emph{IEEE Trans. Wireless Commun.}, vol.~23,
  no.~12, pp. 19\,744--19\,759, Dec. 2024.

\bibitem{ref28}
M.~Olyaee and S.~Buzzi, ``User-centric cell-free massive {MIMO} with access
  points empowered by fluid antennas,'' in \emph{Proc. IEEE Int. Workshop
  Signal Process. Adv. Wireless Commun. (SPAWC)}, Sep. 2024, pp. 666--670.

\bibitem{ref29}
W.~Ma, L.~Zhu, and R.~Zhang, ``Compressed sensing based channel estimation for
  movable antenna communications,'' \emph{IEEE Commun. Lett.}, vol.~27, no.~10,
  pp. 2747--2751, Oct. 2023.

\bibitem{ref30}
J.~Zhu, G.~Chen, P.~Gao, P.~Xiao, Z.~Lin, and A.~Quddus, ``Index modulation for
  fluid antenna-assisted {MIMO} communications: System design and performance
  analysis,'' \emph{IEEE Trans. Wireless Commun.}, vol.~23, no.~8, pp.
  9701--9713, Aug. 2024.

\bibitem{ref31}
X.~Chen, B.~Feng, Y.~Wu, D.~W.~K. Ng, and R.~Schober, ``Joint beamforming and
  antenna movement design for moveable antenna systems based on statistical
  {CSI},'' in \emph{Proc. GLOBECOM IEEE Global Commun. Conf.}, 2023, pp.
  4387--4392.

\bibitem{ref32}
Y.~Ye, L.~You, J.~Wang, H.~Xu, K.-K. Wong, and X.~Gao, ``Fluid antenna-assisted
  {MIMO} transmission exploiting statistical csi,'' \emph{IEEE Commun. Lett.},
  vol.~28, no.~1, pp. 223--227, Jan. 2024.

\bibitem{ref33}
Z.~Zheng, Q.~Wu, W.~Chen, and G.~Hu, ``Two-timescale design for movable
  antennas enabled-multiuser {MIMO} systems,'' 2024, \textit{arXiv:2410.05912}.

\bibitem{ref34}
X.~Zeng, J.~Fang, B.~Wang, B.~Ning, and H.~Li, ``Csi-free position optimization
  for movable antenna communication systems: A derivative-free optimization
  approach,'' \emph{IEEE Wireless Commun. Lett.}, vol.~14, no.~1, pp. 53--57,
  Jan. 2025.

\bibitem{ref35}
T.~X. Doan, C.~D. Ho, and H.~Q. Ngo, ``Massive {MIMO} under multi-keyhole
  channels: Does the use-and-then-forget bounding technique work?'' \emph{Phys.
  Commun.}, vol.~47, p. 101384, Aug. 2021.

\bibitem{ref37}
P.~Gu, S.~Tian, and Y.~Chen, ``Iterative learning control based on nesterov
  accelerated gradient method,'' \emph{IEEE Access}, vol.~7, pp.
  115\,836--115\,842, 2019.

\bibitem{ref36}
L.~Xingsi, ``An entropy-based aggregate method for minimax optimization,''
  \emph{Eng. Optim.}, vol.~18, no.~4, pp. 277--285, Aug. 1992.

\bibitem{ref38}
S.~M. Kay, \emph{Fundamentals of statistical signal processing: estimation
  theory}.\hskip 1em plus 0.5em minus 0.4em\relax Prentice-Hall, Inc., 1993.

\end{thebibliography}


	
	

\vfill

\end{document}